\documentclass{emulateapj}
\usepackage{psfig}
\usepackage{subfigure}
\usepackage{lscape}

\usepackage{verbatim}

\newcommand{\etal}{{et~al.}}

\newcommand\bootes{Bo\"{o}tes}

\bibpunct[;]{(}{)}{;}{a}{}{,}
\def\gtsima{$\; \buildrel > \over \sim \;$}
\def\gsim{\lower.7ex\hbox{\gtsima}}

\usepackage[usenames]{color}

\shorttitle{Strong PAH Emission $\lowercase{z} \approx 2$ ULIRGs}
\shortauthors{Desai \etal}

\begin{document}

\title{Strong PAH Emission from $\mathbf{\lowercase{z} \approx 2}$ ULIRGs\altaffilmark{1}}

\author{Vandana~Desai\altaffilmark{2,3}, 
B.~T.~Soifer\altaffilmark{2,3}, 
Arjun~Dey\altaffilmark{4}, 
Emeric~Le~Floc'h\altaffilmark{5}, 
Lee~Armus\altaffilmark{3}, 
Kate~Brand\altaffilmark{4,6}, 
Michael~J.~I.~Brown\altaffilmark{7}, 
Mark Brodwin\altaffilmark{4,8,9},
Buell~T.~Jannuzi\altaffilmark{4}, 
James~R.~Houck\altaffilmark{10}, 
Daniel~W.~Weedman\altaffilmark{10},
Matthew L. N. Ashby\altaffilmark{8},
Anthony Gonzalez\altaffilmark{11},
Jiasheng Huang\altaffilmark{8},
Howard A. Smith\altaffilmark{8},
Harry~Teplitz\altaffilmark{3}, 
Steve P. Willner\altaffilmark{8},
Jason Melbourne\altaffilmark{2}}

\altaffiltext{1}{Based on observations made with the \textit{Spitzer
Space Telescope}, operated by the Jet Propulsion Laboratory under NASA
contract 1407.}

\altaffiltext{2}{Division of Physics, Mathematics and Astronomy,
California Institute of Technology, Pasadena CA 91125}

\altaffiltext{3}{Spitzer Science Center, California Institute of
Technology, Pasadena CA 91125}

\altaffiltext{4}{National Optical Astronomy Observatory, Tucson AZ
85726-6732}

\altaffiltext{5}{Spitzer Fellow; Institute for Astronomy, University of Hawaii,
Honolulu HI 96822}

\altaffiltext{6}{Space Telescope Science Institute, Baltimore MD
21218}

\altaffiltext{7}{School of Physics, Monash University, Clayton,
Victoria 3800, Australia}

\altaffiltext{8}{Harvard-Smithsonian Center for Astrophysics, 60 Garden Street, Cambridge MA 02138}

\altaffiltext{9}{W. M. Keck Postdoctoral Fellow at the Harvard-Smithsonian Center for Astrophysics}

\altaffiltext{10}{Astronomy Department, Cornell University, Ithaca NY
14853}

\altaffiltext{11}{Department of Astronomy, University of Florida, Gainesville FL 32611-2055}

\begin{abstract}

Using the Infrared Spectrograph on board the \textit{Spitzer Space
  Telescope}, we present low-resolution ($64 < \lambda / \delta\lambda
< 124$), mid-infrared (20--38~$\micron$) spectra of 23 high-redshift
ULIRGs detected in the \bootes \ field of the NOAO Deep Wide-Field
Survey.  All of the sources were selected to have $1)~f_{\nu}(24
\micron) > 0.5~ \rm{mJy}$; 2) $R-[24] > 14$ Vega mag; and 3) a
prominent rest-frame 1.6 $\micron$ stellar photospheric feature
redshifted into \textit{Spitzer's} 3--8~$\micron$ IRAC bands.  Of
these, 20 show emission from polycyclic aromatic hydrocarbons (PAHs),
usually interpreted as signatures of star formation.  The PAH features
indicate redshifts in the range $1.5 < z < 3.0$, with a mean of
$\langle z \rangle = 1.96$ and a dispersion of 0.30.  Based on local
templates, these sources have extremely large infrared luminosities,
comparable to that of submillimeter galaxies.  Our results confirm
previous indications that the rest-frame 1.6 $\micron$ stellar bump
can be efficiently used to select highly obscured starforming galaxies
at $z \approx 2$, and that the fraction of starburst-dominated ULIRGs
increases to faint 24 $\micron$ flux densities.  Using local
templates, we find that the observed narrow redshift distribution is
due to the fact that the 24 $\micron$ detectability of PAH-rich
sources peaks sharply at $z = 1.9$.  We can analogously explain the
broader redshift distribution of \textit{Spitzer}-detected
AGN-dominated ULIRGs based on the shapes of their SEDs.  Finally, we
conclude that $z \approx 2$ sources with a detectable 1.6 $\micron$
stellar opacity feature lack sufficient AGN emission to veil the 7.7
$\micron$ PAH band.

\end{abstract}


\section{Introduction}
\label{sec:intro}

The launch of the \textit{Spitzer Space Telescope} \citep{Werner04}
has allowed the identification and study of significant populations of
distant, infrared-bright galaxies.  The most extreme of these are
exceptionally faint in the optical but readily detected in 24
$\micron$ surveys carried out with \textit{Spitzer}/MIPS
\citep{Rieke04}.  For example, \citet{Dey08} selected a population of
Dust-Obscured Galaxies (DOGs) from the \bootes \ field of the NOAO
Deep Wide-Field Survey \citep[NDWFS;][]{Jannuzi99} via the criteria
$R-[24]>14$~Vega~mag and $f_{\nu}(24 \micron) > 0.3$~mJy \citep[see
also][]{Houck05,Fiore08}.  Drawing on \textit{Spitzer}/IRS
\citep{Houck04} redshifts \citep[this work and][]{Houck05},
near-infrared groundbased spectra \citep{Brand07a}, and optical
spectra \citep{Desai08a}, \citet{Dey08} find that DOGs have a broad
redshift distribution centered at $\langle z \rangle = 1.99$.  These
redshifts imply enormous luminosities, similar to and even exceeding
those of local ULIRGs and distant submillimeter galaxies.  While DOGs
are rare ($\approx$2600 are found over the $\approx$9 square degrees
of the \bootes \ field), \citet{Dey08} estimate that they contribute a
quarter of the infrared luminosity density at $z=2$.  In addition,
their space densities \citep{Dey08} and clustering properties
\citep{Brodwin08} have led to the suggestion that DOGs may be
evolutionarily related to both coeval SMGs and the most massive local
galaxies.

However, the properties of DOGs must be studied more thoroughly before
any possible relationships to other populations can be firmly
established.  One of the most basic unresolved issues is whether the
enormous luminosities of DOGs are powered predominantly by star
formation or AGN activity.  The optical through mid-infrared SEDs
\citep{Dey08} of the DOGs that are brightest at 24 $\micron$ tend to
resemble power-laws, as expected for AGN-dominated sources.  In
contrast, the fainter DOGs include a larger fraction of sources
featuring a bump at rest-frame 1.6 $\micron$.  This bump is
characteristic of old stellar populations.  Its detectability
indicates limited AGN activity because an AGN would result in extra
flux at rest-frame $\approx$2.5 $\micron$, thereby masking the bump.

Follow-up data for the bright ($f_{\nu}(24 \micron) > 0.75$~mJy)
power-law DOGs support the interpretation that they are AGN-dominated.
IRS spectroscopy reveals absorbed power-laws in the mid-infrared, as
expected for AGN \citep{Houck05,Weedman06}.  Near-infrared
spectroscopy of a small sample of bright power-law DOGs indicates that
they host powerful AGN \citep{Brand06}.  Similarly, the far-infrared
SEDs of bright DOGs are similar to the AGN-dominated local ULIRG Mrk
231 \citep{Tyler09}. The morphologies of bright power-law DOGs have
also been examined \citep{Melbourne09, Melbourne08, Bussmann09,
  Dasyra08}, and they tend to be more compact than their fainter
counterparts, as expected for AGN versus starforming regions.

Because they are relatively more difficult to observe, there has been
less follow-up of the faint ($f_{\nu}(24 \micron) < 0.75$~mJy) bump
DOG population.  A stacking analysis shows that their far-infrared
properties are consistent with star-formation, and similar to SMGs
\citep{Pope08}.  However, X-ray stacking analyses have provided mixed
results \citep{Fiore08,Pope08}.  Not many IRS spectra of faint bump
DOGs exist. \citet{Pope08} found that 12 out of 70 DOGs in the GOODs
field have serendipitous IRS spectra.  The sources lie in the range
$0.2 < f_{\nu}(24 \micron) / {\rm mJy} < 1.5$. Of these, half show PAH
features with equivalent widths suggesting that they are dominated by
star formation in the mid-infrared.  The PAH-rich sources tend to be
the fainter ones ($f_{\nu}(24 \micron) < 0.7$ mJy) and have IRAC SEDs
that deviate from a power law.  In addition, IRS spectra exist for
infrared bright galaxies with a range of 24~$\micron$ flux densities
that display the 1.6 $\micron$ stellar bump but do not necessarily
meet the DOG criterion because they are too bright in the optical
\citep{Huang09,Farrah08,Yan07}.  These spectra indicate that the
presence of the bump is correlated to the presence of PAH features in
the IRS spectrum, suggesting that bump sources have mid-infrared
emission dominated by star formation.

Given the potential importance of the DOG population in general, and
the fact that the DOG population grows with decreasing 24 $\micron$
flux density, detailed study of the faint bump DOGs is necessary.  The
first step in this process is building a statistically significant
sample of bump DOGs with redshifts for follow-up study (for example,
with Herschel).  Here we present IRS spectra for an additional 23
faint bump DOGs.  The 20 of these for which we were able to determine
redshifts make up $\approx$23\% of the 86 sources used in the redshift
distribution of \citet{Dey08}.

This paper is organized as follows.  In \S\ref{sec:data}, we describe
our observational data and selection critera.  In \S\ref{sec:Results}
we present our results, namely the redshifts of our targets, the
composite IRS spectrum, and their bolometric luminosities.  We discuss
these results in \S\ref{sec:Discussion} and summarize in
\S\ref{sec:Summary}.  In the following, we use ${\rm H}_0 =
70$~km~s$^{-1}$~Mpc$^{-1}$, $\Omega_{\rm m} = 0.3$, and $\Omega_\Lambda =
0.7$.

\section{Observational Data and Target Selection}
\label{sec:data}
We targeted galaxies for mid-infrared IRS spectroscopy based on
multiwavelength imaging of the 9.3 deg$^2$ \bootes \ field of the
NDWFS.  In the following subsections, we describe the survey data, our
strategy for selecting high-redshift starforming ULIRGs, the IRS
spectroscopy of these ULIRGs, and follow-up observations at 70 and 160
$\micron$.

\subsection{Multiwavelength Imaging of the \bootes \ Field of the NOAO Deep Wide-Field Survey}
\label{sec:SurveyData}

The 9.3 deg$^2$ \bootes \ field of the NDWFS\footnote{See
  http://www.noao.edu/noao/noaodeep/ for more information regarding
  the depth and coverage of the NDWFS.} has been imaged in the $B_W$,
$R$, $I$, and $K$ bands down to 5$\sigma$ point-source depths of
$\approx$27.1, 26.1, 25.4, and 19.0 Vega mags, respectively.
Additional imaging at the $J$ and $K_s$ bands was obtained for 4.7
deg$^2$ of the \bootes \ field through the FLAMEX survey
\citep{Elston06}.  Approximately 8.5 deg$^2$ of the \bootes \ field
has also been mapped (PID 30) with the Infrared Array Camera
\citep[IRAC;][]{Fazio04} on board the \textit{Spitzer Space Telescope}
\citep{Werner04}.  The 5$\sigma$ point-source depths of the IRAC
Shallow Survey are 6.4, 8.8, 51, and 50~$\mu$Jy at 3.6, 4.5, 5.6, and
8~$\micron$, respectively \citep{Eisenhardt04}.  Approximately 8.74
deg$^2$ of the \bootes \ field has been imaged with the Multiband
Imaging Photometer for \textit{Spitzer} \citep[MIPS;][]{Rieke04}.  The
1$\sigma$ point-source depths of the MIPS survey are 0.051, 5, and 18
mJy at 24, 70, and 160~$\micron$, respectively.  Approximately 7
deg$^2$ have been observed at 1.4 GHz with the Westerbork Synthesis
Radio Telescope (WRST).  The data are characterized by a 13\arcsec
$\times$ 27 $\arcsec$ beam and a 1-$\sigma_{RMS}$ limiting sensitivity
of 28 $\mu$Jy \citep{deVries02}.  Although both a reduced mosaic and a
catalog have been made publically
available\footnote{http://wwww.astron.nl./wow/testcode.php?survey=5},
we performed our own photometry on the reduced mosaic to ensure proper
deblending for each IRS target.

\subsection{Selection of high-redshift starforming ULIRGs}
\label{sec:SelectionCriteria}

%24 micron flux density cut based on IRS sensitivity limits.
Our goal was to select starforming ULIRGs at $z \approx 2$ for
follow-up mid-infrared spectroscopy with the IRS.  Based on previous
mid-infrared, near-infrared, and optical spectroscopy, we have
established that a selection criterion of $R-[24] > 14$ Vega mag
results in sources at $z \approx 2$
\citep{Dey08,Desai08a,Brand07,Houck05}.  Given both the sensitivity
limits of the IRS and our desire to select high-luminosity sources, we
only considered objects satisfying $f_{\nu}(24\micron) \ge 0.5$~mJy.
Finally, we chose sources that display a rest-frame 1.6 $\micron$
stellar photospheric feature in their IRAC SEDs.  A prominent 1.6
$\micron$ bump suggests a limited contribution from AGN-heated hot
dust, implying that the luminosity of the source is primarily
generated by star formation.  In order to choose the best candidates,
we eliminated sources that were undetected in any of the IRAC
channels.

%Describe simple SED fitting and show SEDs
We adopted Arp~220 as a template for identifying sources with a
prominent 1.6 $\micron$ stellar bump.  We varied only the template
normalization and redshift in order to fit our Arp~220 template to the
four IRAC flux densities of our sample sources.  We also fit a
power-law to these flux densities.  If the best power-law model
provided a better fit than the best Arp~220 model, we rejected the
candidate.  If the best-fitting Arp~220 model corresponded to a
photometric redshift less than 1.5, we also rejected the source.

For the remaining sources, we tested the dependence of the best-fit
photometric redshift on photometry errors by performing Monte Carlo
simulations.  For each of 500 trials, we randomly perturbed the four
observed IRAC flux densities.  The sizes of the perturbations were
based on Gaussian distributions centered on the observed flux
densities and with dispersions equal to the reported 1$\sigma$
photometric errors.  In this way, we generated an approximate
photometric redshift probability distribution for each source.

The full SEDs and photometric redshift probability distributions for
each source were visually inspected.  Based on these, a total of 23
sources were selected for follow-up IRS spectroscopy.  Their positions
are listed in Table \ref{table:targets}, their photometric
properties are summarized in Table \ref{table:photometry}, and their SEDs
are shown in Figure \ref{fig:SEDs}.  The first fifteen sources were
chosen based upon the tightness of their redshift probability
distributions (favoring sources with the highest signal-to-noise
photometry).  The last eight sources had broad photometric redshift
probability distributions, but were chosen because they had a
significant probability of lying at $z > 2.5$.

\citet{Dey08} find that the DOG selection picks out about 50\% of the
ULIRGs at $z\approx 2$.  We now explore how our present sample of bump
DOGs compares to other $z \approx 2$ galaxies that have been selected
for follow-up with the IRS.  These include sources detected in \bootes
\ \citep{Houck05, Weedman06}, the First Look Survey
\citep[FLS;][]{Yan05, Sajina07} and SWIRE \citep{Farrah08,
  Lonsdale08}.  The \bootes \ IRS targets cited above satisfy the same
$R-[24]$ color cut, but are brighter
($f_{\nu}(24~\micron)~>~0.75$~mJy) than the current sample.  Although
they were selected without regard to IRAC SED shape, the bright 24
$\micron$ flux density cut resulted in the vast majority having
power-law SEDs through the IRAC bands.  The FLS sample consists of
bright ($f_{\nu}(24 \micron) > 0.9$~mJy) 24 $\micron$ sources with
less extreme infrared-to-optical flux density ratios than we imposed.
Also, the FLS sample satisfied a cut of $\log_{10}(\nu S_{24} / \nu
S_{8}) \gsim 0.5$ to try to select for starforming, rather than
AGN-like, sources.  The SWIRE sample was selected to have $f_{\nu}(24
\micron) > 0.5$~mJy and strong 1.6~$\micron$ stellar bumps peaking in
IRAC channel 2 (``bump-2'' sources).  Figure \ref{fig:yancolors} shows
$\log_{10}(\nu S_{24} / \nu S_{8})$ versus $\log_{10}(\nu S_{24} / \nu
S_{R})$ for these different samples.  Our sample has similar values of
$\log_{10}(\nu S_{24} / \nu S_{8})$ compared to the FLS and SWIRE
samples.  This is not surprising since all three programs were
designed to select starforming ULIRGs (see also \citet{Brand06}).
However, the brighter \bootes \ samples of \citet{Houck05} and
\citet{Weedman06} extend to lower values of $\log_{10}(\nu S_{24} /
\nu S_{8})$.  This is consistent with their IRS spectra being
primarily AGN-like.

IRAC color-color diagrams have been used by various groups
\citep[e.g.][]{Lacy04,Stern05,Sajina05} to determine the origin of the
mid-infrared luminosity of galaxies.  In particular, these groups have
defined a wedge-shaped region of this diagram that is populated
predominantly by AGN.  Figure \ref{fig:irac} shows one version of the
IRAC color-color diagram and the AGN wedge, along with data from the
current sample and for several comparison samples.  Most of the
previously-published \bootes \ sources have AGN-like IRAC colors,
consistent with their AGN-like IRS spectra.  The FLS sample includes
some sources in the AGN wedge, but has a higher fraction of sources
that are blue in $S_{8.0} / S_{4.5}$.  The former tend to have
AGN-like IRS spectra, and the latter tend to have PAH-rich IRS
spectra.  Our sample also has very blue $S_{8.0} / S_{4.5}$ colors.
It occupies a similar color-color space as the ``bump-2'' SWIRE
sources, which have PAH-rich IRS spectra.  This plot indicates that
the ULIRGs with the lowest values of $S_{8.0} / S_{4.5}$ are the most
likely to have mid-infrared luminosities dominated by star formation.
The sources with low values of $S_{8.0} / S_{4.5}$ are those in which
the 8 $\micron$ flux turns over, indicating the presence of a 1.6
$\micron$ stellar bump and a lack of AGN-heated hot dust.  Conversely,
sources with larger values of $S_{8.0} / S_{4.5}$ have an AGN-heated
hot dust component which elevates the observed 8 $\micron$ flux
density.  These sources are more likely to have mid-infrared
luminosities dominated by AGN activity.  Although we did not explicity
include a $S_{8.0} / S_{4.5}$ selection criterion, there are no bump
DOGs with $f_{\nu}(24) > 0.5$ mJy in our 24 $\micron$ catalog that
have $S_{8.0} / S_{4.5}$ colors bluer than those in our sample.  This
is likely due to our bump selection criterion, where the bump is
detected in the 4.5 or 5.8 $\micron$ band.

\subsection{Spectroscopic Observations and Data Reduction}

\label{sec:Observations}

Mid-infrared spectra of the 23 candidates described in \S
\ref{sec:SelectionCriteria} were obtained with the IRS on board
\textit{Spitzer}.  The $z \approx 2$ candidates (the first 15 in Table
\ref{table:targets}) were observed in program 30334 and the higher
redshift candidates (the last eight in Table \ref{table:targets}) were
observed in program 40441.  We observed each of our targets with the
Long-Low 1 (LL1) module, which provides wavelength coverage between
$21-34$ $\micron$.  At the photometric redshifts of our targets, this
allows us to detect at least the 7.7 $\micron$ PAH emission complex,
which is the strongest of the various PAH complexes.

The IRS has two observing modes: staring and mapping.  For faint point
sources like ours, mapping mode is recommended (see the January 2006
\textit{Spitzer} memo entitled \textit{Report on ultradeep IRS
  spectroscopy of faint sources}). We therefore used mapping mode to
take spectra of each of our targets in 4 different positions along the
slit.  We first acquired a nearby, bright 2MASS star, and then offset
to the science target.  We chose the longest available ramp time (120
seconds) to minimize overhead.  We used a parallel step size of 33
arcseconds for 4 roughly equally-spaced steps along the slit.  The
number of cycles per position ranged from 7 to 18, depending on the 24
$\micron$ flux density and the expected redshift of the source.  The
effective exposure times ranged from 853 to 2194 seconds.

The raw data were processed by the S13 pipeline at the \textit{Spitzer
  Science Center}.  The pipeline performed ramp fitting, dark sky
subtraction, droop correction, linearity correction, flat fielding,
and wavelength calibration.  See the IRS Data
Handbook\footnote{http://ssc.spitzer.caltech.edu/irs/dh/} for more
details on these steps.  The resulting data products are known as the
Basic Calibrated Data (BCD).  In the following we describe additional
processing of the two-dimensional spectra (residual charge removal,
rogue pixel interpolation, and sky subtraction) and spectral
extraction.

Although the detector is reset prior to each integration, a small
fraction (1--2\%) of the charge persists between frames.  If no bright
sources are observed, then the zodiacal background will be the major
contributor to the residual charge.  Over long AORs, the residual charge
can build to a significant level, and must be removed.  We subtracted
the background in each BCD by subtracting off the median of the counts
in pixels 25 through 58 (inclusive) of each row.  These pixels were
chosen because they are unaffected by source flux.

%Masking.
We interpolated over unstable (``rogue'') pixels in the
background-subtracted frames.  A mask of known rogue pixels is
provided by the SSC, and we identified further bad pixels from the
data themselves.  We used the IRSCLEAN program provided by the SSC to
find rogue pixels in the 2D data, using the default settings.  We
further searched for pixels with abnormally high variance
($>10\sigma$) with time.  Pixels were also masked if either the
background or the noise surpassed a certain threshhold (1000 for
both).  All masked pixels were interpolated over using IRSCLEAN.

%Background for each position.
Once the 2D images were cleaned of rogue pixels, we created sky images
from the data.  The background for each target position was computed by
taking the median of the cleaned images for all of the other positions.

%Coaddition of frames.
%Combination of spectra.
The individual reduced frames were coadded to produce final 2D spectra
at each map position.  One dimensional spectra were extracted at each
map position using the SPICE software provided by the SSC.  The
optimal extraction option was used.  When this option is set, an
extraction aperture is not set.  Instead, each pixel is weighted by
its position, based on the spatial profile of a bright calibration
star.  After extraction, we are left with 4 1D spectra per source.
These were averaged together to produce the final 1D spectrum.

%Signal-to-noise
The signal-to-noise of the resulting spectra is difficult to quantify
because of the lack of continuum.  We estimate a noise level of
0.1--0.2~mJy by measuring the dispersion in the residuals of each
spectrum after subtracting off the smoothed composite spectrum (see
Section \ref{sec:medianspectrum}).

\subsection{Pointed MIPS 70 and 160 $\micron$ observations and reduction}
\label{sec:MIPS}

Although the multiwavelength survey of the entire \bootes \ field does
include \textit{Spitzer} MIPS 70 and 160 $\micron$ observations, these
are too shallow to provide interesting limits on the far infrared SEDs
of the DOGs.  The 3-$\sigma$ limits are 15 and 54~mJy at 70 and 160
$\micron$, respectively.  We therefore obtained deeper, pointed
observations of these sources (program 30519, PI Le Floc'h).  The data
were reduced using version 3.06 of the MIPS Data Analysis Tool
\citep[DAT;][]{Gordon07}.  None of our sources were detected at either
70 or 160 $\micron$.  In \S{\ref{sec:luminosities}} we show that the
70~$\micron$ non-detections are consistent with expectations based on
low-redshift templates with IRS spectra similar to these
higher-redshift sources.  We used the following procedure for
determining limits at both wavelengths.

For the 70 $\micron$ data, aperture photometry was performed at the
position of the MIPS 24 $\micron$ source, using a 16$\arcsec$ radius
aperture, an 18$\arcsec$ radius inner sky annulus, and a 39$\arcsec$
outer sky annulus.  We applied the aperture correction tabulated in
the MIPS Data Handbook.  We also measured the dispersion
resulting from performing similar measurements over empty areas of the
70 micron images.  The adopted 3$\sigma$ limit (5~mJy) is taken as
three times this dispersion.

For the 160 $\micron$ data, aperture photometry was performed at the
position of the MIPS 24 $\micron$ source, using a 32$\arcsec$ radius
aperture, a 64$\arcsec$ inner sky annulus, and a 128$\arcsec$ outer
sky annulus.  We applied the aperture correction tabulated in the
MIPS Data Handbook.  The limit was calculated in the same way as for
the 70 $\micron$ data.  In this case, the limit turned out to be
30~mJy, but was very uncertain due to the small number of
pixels in the 160 $\micron$ images.  To be conservative, we adopted
the higher limit of 40.8~mJy predicted by the \textit{Spitzer}
SENS-PET.

\section{Results}
\label{sec:Results}

The IRS spectra of our 23 targets are shown in Figure
\ref{fig:individualspectra}.  Strong PAH features are common in this
sample, especially when compared to the brighter ULIRGs observed by
the IRS in \bootes \ \citep{Houck05,Weedman06}.

In this section we first describe how we determined the spectroscopic
redshifts of these sources.  We then use these spectroscopic redshifts
to present the composite spectrum of the sources with measured redshifts.
From the composite spectrum, we measure the characteristic 7.7 $\micron$
PAH equivalent width.  Finally, we estimate the bolometric and
rest-frame 24 $\micron$ luminosities of these sources, so that they
may be easily compared with other samples, both coeval and at low
redshift.

\subsection{Redshifts}
\label{sec:Redshifts}

Many of the spectra displayed in Figure \ref{fig:individualspectra}
contain at least one PAH emission feature (centered at rest-frame 6.2,
7.7, or 11.3 $\micron$) from which we can determine a redshift.
Because these PAH features are fairly broad and the spectra are of
only moderate signal-to-noise, we choose to measure the redshift using
templates rather than fitting the lines individually.  We use four
templates that contain strong PAH features: the unobscured Galactic
reflection nebula NGC 7023, which represents a pure photo-dissociation
region \citep{Werner04ngc7023}; the average starburst spectrum of
\citet{Brandl06}; and the prototypical starbursts NGC 7714
\citep{Brandl04} and M82 \citep{Sturm00}.  Figure \ref{fig:examplez}
shows an example of our redshift determination.  After computing the
best-fit redshift resulting from each template, we adopt the redshift
associated with the template that provides the best formal fit.  In 12
cases, this was NGC 7023; in four cases this was the average starburst
spectrum; in three cases this was NGC 7714; and in one case this was
the M82 template.  We were unable to determine a spectroscopic
redshift for three sources because they had no strong features.  The
adopted redshift for each source is listed in Table
\ref{table:targets}.  For a given source, all of the templates
provided very similar redshifts.  The redshift errors presented in
Table \ref{table:targets} are based on the range of redshifts computed
for a given source.  The average redshift of sources in our sample is
$\langle z \rangle = 1.96$, and the dispersion is 0.30.

Figure \ref{fig:mediansed} presents the rest-frame SEDs, scaled to
unity at 1.6 $\micron$, of all 20 sources with spectroscopic
redshifts.  For comparison, we overplot the SEDs of M82
\citep{Sturm00}, Arp~220 \citep{Armus07}, Mrk~231 \citep{Armus07}.
The SEDs of the $z \approx 2$ starforming ULIRGs most closely resemble
that of Arp~220, in terms of their $R-[24]$ colors.  The SEDs of our
sample appear to be much more strongly reddened than that of M82.
Again, this is unsurprising given the selection criterion $R-[24] >
14$ Vega mag.  Also as expected, the SEDs of the high-redshift ULIRGs
are significantly colder than that of Mrk~231.  The exception is
Source 21 at $z=3.04$.  Its observed-frame 24 $\micron$ (rest-frame
5.9) flux density is more consistent with the Mrk~231 template than
that of Arp~220.
 
\subsection{Composite spectrum}
\label{sec:medianspectrum}

Because the signal-to-noise ratio of any individual spectrum in Figure
\ref{fig:individualspectra} is fairly low, we provide three composite
spectra in Figure \ref{fig:medianspectrum}: the median, the mean, and
the variance-weighted mean.  To compute the composite spectra, we
shifted the wavelengths of each spectrum to the rest-frame, based on
the redshifts calculated in \S{\ref{sec:Redshifts}}; sampled the
rest-frame wavelengths to a common grid; and normalized each spectrum
to unity at 7.7 $\micron$.  

The median spectrum shows strong PAH features at 7.7, 8.6, and 11.3
$\micron$.  In the bottom panel of Figure \ref{fig:medianspectrum}, we
compare this median spectrum to an average starburst spectrum compiled
by \citet{Brandl06}, to a median cold ($S_{25}/S_{60} < 0.2$) ULIRG
spectrum compiled by \citet{Desai07b}, and to the spectrum of Arp~220
\citep{Armus07}.  The wavelength range between 9 and 11 $\micron$ can be
strongly affected by a broad silicate absorption feature centered at
9.7 $\micron$.  Unfortunately, the limited wavelength range of our
spectra prevent an accurate determination of the unabsorbed continuum
level, and we are therefore unable to precisely quantify the depth of
the silicate absorption.  However, Figure \ref{fig:medianspectrum}
suggests that this absorption is comparable to that of local ULIRGs
with cold SEDs.

In Figure \ref{fig:measureew} we compare the median mid-infrared
spectrum of our starforming DOGs to those of other high-redshift
galaxies observed with the IRS (obtained via private communication).
The first comparison sample is a subset of the
\textit{Spitzer}-selected ULIRGs detected in the First-Look Survey
\citep{Yan05,Sajina07}.  Of 48 targets spanning the redshift range $1
< z < 3$, 25\% show strong PAH features.  The average strong-PAH
spectrum is shown in the second panel of Figure \ref{fig:measureew}.

The second comparison sample consists of 24 SMGs (23 detected), with
19 displaying PAH features, in the redshift range $0.65 < z < 3.2$
\citep{MenendezDelmestre09}.  Of these, 19 display prominent PAH
emission.  The composite spectrum from this sample is shown in the
third panel of Figure \ref{fig:measureew}.

The third comparison sample consists of 13 GOODS-N submillimeter
galaxies in the redshift range $0.9 < z < 2.6$ \citep{Pope08}. Of
these, 11 have mid-infrared luminosities dominated by PAH emission.
The composite PAH-rich spectrum from this sample is shown in the
fourth panel of Figure \ref{fig:measureew}.

Although Figure \ref{fig:medianspectrum} provides the visual
impression that the $z \approx 2$ bump DOGs have mid-infrared spectral
properties intermediate between those of local cold ULIRGs and local
starburst galaxies, we wish to measure the equivalent widths of the
PAH emission to facilitate a quantitative comparison to other samples.
Three PAH emission features are clearly seen in Figure
\ref{fig:medianspectrum}, including those centered at rest-frame 7.7,
8.6, and 11.3 $\micron$.  The 11.3 $\micron$ feature is strongly
affected by the rest-frame 9.7 $\micron$ silicate feature.  The other
two features are close enough in wavelength that they are often
measured together.  We adopt this procedure and call the result the
7.7 $\micron$ PAH equivalent width.  Because our spectrum has limited
wavelength coverage, the continuum is quite difficult to determine.
We therefore adopt an empirical method similar to that used by
\citet{Spoon07}.  The technique is illustrated in Figure
\ref{fig:measureew}.  The resulting rest-frame equivalent width for
the 7.7 $\micron$ PAH feature is 0.5$\pm$0.05 $\micron$ for our
sample.  Given the difficulties of this measurement, this is similar
to the values of 0.4$\pm$0.05 $\micron$ that we find for the comparison
samples.  It is also similar to the values found for local starbursts
\citep{Brandl06}.

In addition to constructing a single composite spectrum, we also tried
binning the spectra by observed properties to explore the variation
within our sample.  For example, a trend with spectral type and
luminosity is observed in the local universe (see
\S{\ref{sec:localcomparison}}).  It has also been seen in the $z
\approx 2$ sample of \citet{Sajina07}. However, we find no convincing
correlation between spectral shape and $\nu {\rm L}_{\nu}(24 \micron)$
in our sample (see \S{\ref{sec:luminosities}} for derivation of
luminosity).  In both the low and high redshift samples where a
correlation is seen, the scatter is very large, and can only be seen
when comparing the full range of spectral types (AGN-dominated to
starburst-dominated).  Therefore, the lack of a correlation within our
modest sample of primarily starburst-dominated sources is consistent
with previous findings.  We also explored spectral variations with the
colors plotted in Figures \ref{fig:yancolors} and \ref{fig:irac}, and
found no convincing trends.

\subsection{Luminosities}
\label{sec:luminosities}

Given the redshifts calculated in \S{\ref{sec:Redshifts}} ($\langle z
\rangle = 1.96$), we can estimate the infrared luminosities of our
sources.  Our estimates are necessarily crude because we currently
have no direct flux measurements at wavelengths that probe the
expected peak ($\approx$ rest-frame 70 $\micron$) of the spectral
energy distributions of our sources.  Our longest wavelength
photometric detection comes from \textit{Spitzer} MIPS observations in
the 24 $\micron$ band, which corresponds to a rest-frame of
$\approx$8.4 $\micron$ for these sources.  Although we have 70 and 160
$\micron$ observations (corresponding to rest-frames 24 and 55
$\micron$, respectively), these provide only limits, not detections
(see \S{\ref{sec:MIPS}}).

Given our lack of direct constraints on the far-infrared SEDs of our
sources, we extrapolate from the rest-frame mid-infrared to determine
L$_{8-1000 \micron}$.  There are a variety of bolometric correction
factors calibrated on local sources \citep[e.g.][]{Caputi07} and several
families of templates \citep[e.g.][]{Chary01, Dale05, Siebenmorgen07,
  Rieke09} that can be used to guide this extrapolation.
Unfortunately, the resulting value of L$_{8-1000 \micron}$ can vary by
a factor of 5 depending on which local relation or template is used
\citep{Dale05,Caputi07,Dey08}.  Thus, it is important to choose local
templates that are as similar as possible to the sample with unknown
far-infrared SEDs.

A first pass with the starburst templates of \citet{Chary01} indicates
that the bump DOGs have ULIRG luminosities.  To refine our estimate of
the bolometric luminosity, we therefore wish to use a local ULIRG
sample for our extrapolation.  The \textit{Spitzer} IRS GTO Team
obtained IRS spectra for $\approx$100 local ULIRGs spanning a range of
infrared colors and luminosities \citep{Armus07, Desai07b, Farrah07,
  Spoon07}.  We use this sample to estimate L$_{8-1000 \micron}$ using
the following procedure.  First, we fit each of the 107 GTO ULIRGs to
the median spectrum shown in Figure \ref{fig:medianspectrum}.  We then
ranked the GTO ULIRGs according to how well they fit the median bump
spectrum, as measured by the $\chi^2$.  We fit the first-ranked GTO
ULIRG to each of the individual bump spectra, yielding infrared
luminosities in the range $2.5 < {\rm L}_{8-1000 \micron}/10^{12} {\rm
  L}_{\odot} < 25$.  This variation is due both to the fact that the
bump sources have a variety of 24 $\micron$ flux densities (from 0.5
to 0.75 mJy) as well as a range of redshifts (from 1.5 to 3.0).  To
examine the possible error in the value of ${\rm L}_{8-1000 \micron}$
for a given bump source, we calculated the range in ${\rm L}_{8-1000
  \micron}$ that results from the most highly ranked 10, 25, and 50
local GTO ULIRGs.  These result in ${\rm L}_{8-1000 \micron}$ ranges
that are within a factor of 2, 3, and 3 of the best-fit ${\rm
  L}_{8-1000 \micron}$.  We estimate that the individual values of
${\rm L}_{8-1000 \micron}$ are good to a factor of $\approx$3.  The
results are listed in Table \ref{table:luminosities}.

The radio fluxes and limits presented in Table \ref{table:photometry}
(see also \S{\ref{sec:SurveyData}}) can also be used to constrain
${\rm L}_{8-1000 \micron}$.  At low redshift, a far-infrared-radio
correlation has been observed to hold for a wide variety of galaxy
types over four orders of magnitude in luminosity
\citep[e.g.][]{deJong85,Helou85,Condon92,Yun01,Bell03}.  Such a
correlation has also been observed at higher redshifts \citep[$0.6 < z
< 3$;][]{Gruppioni03,Appleton04,Frayer06a,Kovacs06}.  To illustrate
the systematic uncertainties in using the far-infrared-radio
correlation to estimate ${\rm L}_{8-1000 \micron}$ for our bump
sources, we use two different calibrations of the correlation.  The
first calibration is that of \citet{Bell03}, and is based on a sample
of 164 local galaxies without signs of AGN activity.  The second is by
\citet{Kovacs06}, and is determined from 15 high-redshift ($1 < z <
3$) SMGs.  The high-redshift calibration yields smaller values (by a
factor of $\approx$4) of ${\rm L}_{8-1000 \micron}$ for a measured
radio flux.  It is consistent with the findings of \citet{Murphy08}
for 11 SMGs out to $z = 2.5$.  The infrared luminosities resulting
from each calibration are presented in Table \ref{table:luminosities}.
Despite the large uncertainties, the radio-derived luminosities are
consistent with the interpretation that these sources are ULIRGs.
Because a significant fraction of our sources have only radio limits,
in the following we adopt the values of ${\rm L}_{8-1000 \micron}$
derived by extrapolating the IRS spectra of our sources using local
ULIRG templates (column 2 of Table \ref{table:luminosities}).

If all of this luminosity is produced by star formation, this would
correspond to star formation rates in the range 400--4000 M$_{\odot}$
yr$^{-1}$ \citep{Kennicutt98}. In reality, some fraction of this
infrared luminosity is likely due to AGN activity.  Unfortunately,
given the limited wavelength range of our spectrum, we cannot
disentangle the fraction of the mid-infrared luminosity that is
contributed by star formation versus AGN activity.  However, the AGN
contribution is likely not dominant, since a strong power-law
component would overwhelm both the 1.6 $\micron$ stellar bump and the
PAH features.  Although these estimates are uncertain, they indicate
that the sources we have identified are extremely luminous, comparable
to the coeval submillimeter galaxies.  Another possibility is that the
local templates differ significantly from our high-redshift sources in
the far-infrared.  Direct measurements at long wavelengths are needed
to explore this possibility.

In \S\ref{sec:MIPS}, we described follow-up 70 and 160~$\micron$
follow-up observations of a subset of the sources presented in this
paper.  All were non-detections, with 3-$\sigma$ limits of 5 and
40.8~mJy at 70 and 160~$\micron$, respectively.  If our sources are
similar in luminosity to SMGs, then these limits should be consistent
with the far-infrared SEDs measured for SMGs.  Using the SMG SEDs
presented in Figure 5 of \citet{Pope08}, we estimate that SMGs with
24~$\micron$ flux densities similar to our bump sources should have
observed 70~$\micron$ flux densities of $\sim$2~mJy and 160~$\micron$
flux densities of $\sim$20~mJy.  These estimates are below our
detection limits, and thus our far-infrared non-detections are still
consistent with SMG-like luminosities.

Similarly, the wavelength coverage of the available IRS spectra allows
us to predict the 70 $\micron$ flux density that would be observed if
the local GTO ULIRGs were shifted out to $z \approx 2$.  Using the 10
GTO ULIRGs that provide the best fit to the median spectrum shown in
Figure \ref{fig:medianspectrum}, we estimate $f_{\nu}(70 \micron) =
0.5 - 1$ mJy.  Thus, the observed-frame far-infrared SEDs of our bump
sources are consistent with low-redshift ULIRGs.

\section{Discussion}
\label{sec:Discussion}

\subsection{Why does our selection of starforming ULIRGs lie within such a narrow redshift range?}
\label{sec:selection}

Table \ref{table:targets} shows that our sample of starforming ULIRGs
has a very narrow range of redshifts: $\langle z \rangle = 1.96 \pm
0.3$.  We imposed two selection criteria which impact this
distribution: (1) our requirement that we see the 1.6 $\micron$
stellar bump in the IRAC bands; and (2) our requirement that
$f_{\nu}(24 \micron) > 0.5$ mJy.  In this section, we explore how
these criteria could have resulted in the observed narrow redshift
distribution.

Based on local templates, is it plausible that requirement (1) results
in the observed narrow redshift range?  In Figure \ref{fig:irac}, we
showed that requirement (1) translates roughly into selecting targets
from a region of IRAC color-color space defined by $\log_{10}(S_{8.0}
/ S_{4.5}) < 0.2$ and $0.0 < \log_{10}(S_{5.8} / S_{3.6}) < 0.3$.  In
the top two panels of Figure \ref{fig:selection}, we plot the expected
IRAC colors of three local templates as a function of redshift.  The
three templates span a range of properties: Arp~220 is PAH-rich ULIRG,
08572+3915 is a deeply obscured AGN-dominated ULIRG, and the starburst
template is an average of lower-luminosity starbursts from
\citet{Brandl06}.  Figure \ref{fig:selection} shows that Arp~220 meets
requirement (1) over the redshift range $z > 1.75$, while 08572+3915
and local starbursts would never be identified as bump sources over
the redshift range plotted.  We conclude that our requirement of
seeing the 1.6 $\micron$ bump in the IRAC bands selects for high
redshift sources ($z \gsim 1.75$), but can not by itself explain the
narrowness of the resulting redshift distribution.

To investigate how requirement (2) affects the redshift range of our
sources, we also show the expected 24 $\micron$ flux density of our
three templates as a function of redshift in the bottom panel of
Figure \ref{fig:selection}.  All templates show a dip at $z \approx
1.5$.  One cause for this dip is that the 24~$\micron$ bandpass is
sampling the 9.7 $\micron$ silicate absorption feature at $z \approx
1.5$.  In Arp~220 and in the starburst template, an additional reason
for this dip is that the 24 $\micron$ bandpass samples the 8.6 and
11.3 $\micron$ PAH emission features at both higher and lower
redshifts, respectively.  In these two sources, the 7.7 and 8.6
$\micron$ PAH emission features passing through the 24 $\micron$
passband cause a peak at $z = 1.8-1.9$.  The 24 $\micron$ flux
density of the AGN-dominated ULIRG 08572+3915 does not fall as sharply
at $z > 1.9$ because of strong continuum emission longward of
rest-frame 8.6 $\micron$.  Also shown in Figure \ref{fig:selection} is
the redshift distribution of the bump DOGs with IRS redshifts.  The
redshifts of the bump sources coincide with the peak seen in the Arp
220 template.  This coincidence suggests that the narrow redshift
distribution of the bump sources is due to the fact that they are
preferentially detected over the redshift range where their PAH
emission features are sampled in the 24 $\micron$ bandpass.

Figure \ref{fig:selection} formally allows for the possibility of a
significant population of bump DOGs at $z > 2.2$ that are missed by
our selection technique because there is no PAH feature in the 24
$\micron$ bandpass to boost the 24 $\micron$ flux density enough to
meet our flux cut.  However, there is a second reason why we do not
see bump DOGs at $z > 2.2$.  On average, bump sources that satisfy our
24 $\micron$ detection limit at $z=2.2$ ($z=2.5$) would have to be
more luminous by a factor of $>$1.25 ($>$2) compared to those at $z =
2$.  Since the top end of the galaxy luminosity function typically
declines exponentially, this could be the reason that we do not see
exceptionally luminous DOGs powered by star formation at $z > 2.2$.
Thus, both the mid-infrared spectral shape of starforming ULIRGs and
the declining luminosity function at high luminosities are responsible
for the decline in the redshift distribution at $z > 2.2$ (and the
resulting tightness of the redshift distribution).  Which mechanism
dominates is unclear from the data in hand.

In the bottom panel of Figure \ref{fig:selection} we also show the
redshift distribution of the bright ($f_{\nu}(24 \micron) > 0.75$ mJy)
power-law DOGs from \citet{Houck05}.  For $1.3 < z < 2.9$, this
distribution tracks the profile of 08572+3915.  At lower redshifts,
sources become brighter in the $R$-band, and fail to make our $R-[24]$
DOG cut.  This suggests that the redshift distribution of power-law
DOGs is determined in large part by the SED shape of AGN-dominated
ULIRGs.  The fact that power-law DOGs are found at greater redshifts
than bump DOGs suggests that the luminosity function of power-law DOGs
may be shifted to higher luminosities compared to bump DOGs.  This is
consistent with the observation that the most luminous ULIRGs in the
local universe tend to be AGN-dominated
\citep[e.g.][]{Veilleux99,Tran01,Desai07b}.

The presence of a bump in the IRAC SEDs allowed the selection of
high-redshift ($z \ge 1$) sources, but the additional requirement of a
24 $\micron$ detection tightly constrained the redshift distribution
to $z \approx 2$, for the reasons described above.  For example,
\citet{Brodwin06} use the optical, near-infrared, and IRAC data
available in the \bootes \ field to compute photometric redshifts for
a sample selected at 4.5 $\micron$.  They do not use the 24 $\micron$
flux densities.  Using a hybrid method involving both neural net and
template-fitting techniques (but no ULIRG templates), calibrated using
over 15,000 spectra, they achieve an accuracy of $\sigma = 0.06(1 +
z)$ for 95\% of galaxies at $0 < z < 1.5$ and $\sigma = 0.12(1 + z)$
for 95\% of AGN at $0 < z < 3$.  However, the photometric redshifts
computed for the sample presented here range from $1 < z_{\rm phot} <
3.5$, with an accuracy of $\sigma = 0.8$.  This shows that
photometric redshifts that achieve high accuracy for the majority of
objects do not perform nearly as well on these rare, extreme DOGs.

\subsection{Comparison to Local ULIRGs}
\label{sec:localcomparison}

We intitially used Arp~220 as a template for selecting bump DOGs.
Figure \ref{fig:mediansed} illustrates that the SEDs of bump DOGs are
consistent with that of Arp~220.  However, Figure \ref{fig:selection}
shows that neither Arp~220 nor 08572+3915, if placed at $z=2$, would
be selected in this study.  The reason for this is that they would be
too faint at 24 $\micron$, by factors of 25 and 2, respectively.  Even
if we scaled Arp 220 to match the observed 4.5 $\micron$ flux
densities of our bump DOGs at $z = 2$, it would not make our 24
$\micron$ cut.  In contrast, a scaled Mrk~231 at $z = 2$ would.  This
implies that PAH features are not necessary to boost the
observed-frame 24 $\micron$ flux density of a given source into our
sample; continuum is sufficient.  The fact that we see very few
continuum-dominated sources in our bump sample implies that $z \approx
2$ galaxies with only a small AGN contribution at observed 8 $\micron$
also have a limited AGN contribution at observed-frame 24 $\micron$.

Figure \ref{fig:comparison} shows the rest-frame equivalent widths of
the 7.7 and 11.3 $\micron$ PAH features as a function of rest-frame 24
$\micron$ luminosity for the bump sample, the $z\approx2$ PAH-rich
\textit{Spitzer}-selected FLS sample from \citet{Sajina07}, and for
the $\approx$100 local ULIRGs from \citet{Desai07b}.  This plot
illustrates that the population of high-redshift starbursts that
\textit{Spitzer} is detecting includes galaxies that are uncommonly
luminous at rest-frame 24 $\micron$ compared to ULIRGs in the local
universe.

\subsection{Why do bump sources have strong PAHs?}

The existing IRS spectra of ULIRGs at $z \approx 2$ indicate that
prominent rest frame 1.6 $\micron$ stellar photospheric features are
accompanied by large PAH equivalent widths.  The corollary also
holds: 24 $\micron$ sources with power-law rest frame near-infrared
SEDs have small PAH equivalent widths.  

This trend can be understood within a commonly accepted framework,
wherein the rest-frame near-infrared and mid-infrared spectrum of a
galaxy is a combination of three basic components: stars, the ISM, and
an AGN \citep[e.g.][]{Huang07}. In galaxies selected by their $R-[24]$
colors, the SEDs of these components are also strongly affected by
dust attenuation.  Old stars have an SED resembling a blackbody that
peaks at 1.6 $\micron$ in the rest frame and falls off steeply towards
longer wavelengths.  In contrast, dust-obscured young stars have an
infrared SED which is characterized by mid-infrared PAH features with
high equivalent widths and a broadband SED which rises steeply beyond 6
$\micron$.  Finally, the AGN contributes a power-law SED through the
near- and mid-infared.

Within this framework, galaxies with an observable 1.6 $\micron$
stellar bump have a limited AGN contribution.  In contrast, galaxies
without an observable 1.6 $\micron$ bump have an AGN contribution
which is large enough to disguise the underlying stellar emission by
providing additional flux on its long-wavelength side (rest frame
$\approx$2.5 $\micron$).  If an AGN is emitting significantly at such
short wavelengths, and if it has a power-law SED, then the expected
flux at rest frame wavelengths of $\approx$8 $\micron$ would be even
more significant.  A strong AGN-heated dust component at 8 $\micron$
would increase the continuum under the PAH features.  The result is a
low equivalent width for PAH features in sources with weak 1.6
$\micron$ stellar bumps.  Similarly, a source which does not have
enough AGN-heated dust to swamp out the 1.6 $\micron$ stellar bump
will not have enough AGN-heated dust to depress the equivalent width
of the mid-infrared PAH features.  In this scenario, the PAH features
are \textit{diluted} by AGN-heated smooth dust continuum.  The degree
of the dilution depends on the luminosity and extinction of the AGN
relative to that of the starburst component.  \citet{Teplitz07} mix
the SEDs of a local starburst (NGC 7714) and a pure QSO (PG 0804+761)
to quantitatively explore how an AGN contribution affects both the 1.6
$\micron$ stellar bump and the PAH features.  They find that a weak
AGN contributing 10\% of the total 1--1000 $\micron$ luminosity of a
composite source can hide the 1.6 $\micron$ bump and decrease the PAH
EQW by over 40\%.

While it is generally agreed upon that weak PAH features indicate a
strong AGN component, PAH dilution is not the only scenario that has
been invoked as an explanation.  It is also possible that PAH carriers
are \textit{destroyed} in the harsh radiation fields produced by AGN
\citep{Aitken85,Voit92}.  With the data in hand, we cannot distinguish
between the dilution and destruction scenarios within the $z\approx2$
bump sources.

There is some evidence that PAH dilution occurs in local ULIRGs.
Among local cold ULIRGs, the median 11.3 $\micron$ PAH equivalent
width (EW) is 80\% of that found in starburst galaxies, while the
median 6.2 $\micron$ PAH EW is only 50\% \citep{Desai07a}.  In the
destruction scenario, the EWs of both PAH features should be reduced.
However, the observation that the 11.3 $\micron$ PAH EW is more
similar to the starburst value finds a natural explanation in the
dilution scenario.  Namely, the emission from the hot dust responsible
for the dilution at 6.2 $\micron$ is highly extincted at 11.3
$\micron$, due to the strong, broad absorption feature centered at 9.7
$\micron$.  Whether or not these results based on lower redshift
ULIRGs apply to higher redshift samples remains an open question.

\section{Conclusions}
\label{sec:Summary}

We have tested a technique for using mid-infrared photometry to select
strongly starforming and highly obscured ULIRGs at $z \approx 2$.  In
a previous work, we demonstrated that the simple criteria $R-[24] >
14$~Vega mag and $f_{\nu}(24 \micron) > 0.3$ mJy select for about half of the
ULIRGs at $z=2$ \citep{Dey08}.  We refer to the resulting sample as
Dust-Obscured Galaxies, or DOGs.  Initial studies of the DOGs with the
brightest 24~$\micron$ flux densities ($f_{\nu}(24 \micron) >
0.75$~mJy) revealed AGN-dominated sources.  In this paper, we
attempted to select the most vigorously star-forming DOGs.  Our
technique was to search for DOGs whose IRAC (3--8~$\micron$) SEDs
show evidence for the redshifted 1.6 $\micron$ stellar photospheric
bump characteristic of old stars.  Sources with strong AGN components
should have enough hot dust to swamp this feature.  Thus, sources with
clear 1.6 $\micron$ bumps should have a limited AGN contribution to
their mid-infrared luminosities.  These sources are likely to be
dominated by star formation.  In addition to selecting for starforming
galaxies, this technique also selects for galaxies at $z \approx 2$,
since the 1.6 $\micron$ bump must be shifted into IRAC channels 2 or 3
(4.5 or 5.8 $\micron$, respectively) in order to be easily detected.
To test this selection, we obtained mid-infrared IRS spectroscopy of
23 candidate star-forming galaxies.  Based on these data, we come to
the following conclusions:

\begin{enumerate}

\item Of 23 targets, 20 have mid-infrared spectra displaying PAH
  emission lines indicating that 1) these sources lie at $\langle z
  \rangle = 1.96$ with a dispersion of 0.3; and 2) these sources are
  strongly starforming.  These results indicate that even for the most
  obscured galaxies, the rest-frame 1.6 $\micron$ stellar photospheric
  bump can be used to efficiently select starforming galaxies at $z
  \approx 2$.  This redshift distribution implies that the space
  density of bump DOGs ($R-[24] > 14$~Vega~mag, $0.0 <
  \log_{10}(S_{5.8}/S_{3.6}) < 0.3$, $-0.3 < \log_{10}(S_{8.0} /
  S_{4.5}) < 0.2$) is $\approx$7.5~$\times$~10$^{-5}$ Mpc$^{-3}$,
  consistent with the space density of SMGs in GOODS-N
  \citep[private communication;][]{Wall08}.

\item We have confirmed that DOGs with faint 24~$\micron$ flux
  densities are more likely to be starforming than brighter DOGs.
  \citet{Brand06} used X-ray data to show that this is true for the
  general population of 24 $\micron$ sources.  \citet{Dey08} showed
  that the fraction of DOGs with AGN-like (power-law) IRAC SEDs
  smoothly decreases with decreasing 24 $\micron$ flux density.  In
  this paper, we verify that DOGs which show evidence of the 1.6
  $\micron$ stellar opacity feature in their mid-infrared photometry
  also show strong PAH emission, and appear to be powered principally
  by star formation.

\item Local templates predict that our sample of starforming DOGs have
  infrared luminosities in the range ${\rm L}_{8-1000 \micron} = $ (6--50)
  $\times 10^{12}$ L$_{\odot}$.  If all of this luminosity is due to
  star formation, the star formation rates among our sample would be
  1000--8500~M$_{\odot}$~yr$^{-1}$.  AGN may be contributing to the
  total infrared luminosity, but the strong PAHs suggest that star
  formation dominates at least the mid-infrared luminosity.  The
  unphysically large star formation rates implied may point to an
  evolution in the template from high to low redshift.  This is also
  implied by the fact that starburst galaxies as luminous as those
  recently detectable by \textit{Spitzer} are uncommon in the local
  universe (\ref{fig:comparison}).  Other authors have found hints
  that this may be the case \citep[e.g.][]{Desai07a,Rigby08}.  While
  we have limits on the rest-frame 24 and 55 $\micron$ flux densities
  of these DOGs from MIPS 70 and 160 $\micron$ data, direct
  measurements of the far-infrared SEDs of high-redshift dusty
  galaxies are critical to determining their contribution to the
  global star formation rate at that epoch.

\item The redshift distribution of bump DOGs is remarkably narrow
  because at $z = 1.9$, the strong 7.7 $\micron$ PAH feature boosts
  the 24 $\micron$ flux, pushing sources with insufficient continuum
  into our flux-limited sample.  Analogously, the redshift
  distribution of power-law DOGs is dictated by the fact that sources
  with sufficient continuum to meet the 24 $\micron$ flux density cut
  fall out of the sample at $z \approx 1.5$, when the 9.7 $\micron$
  absorption feature is in the 24 $\micron$ bandpass.  

\item The dearth of objects with the 1.6 $\micron$ stellar feature
  with AGN-like mid-infrared spectra indicates that objects lacking
  enough AGN emission at 2 $\micron$ to hide the stellar bump also
  lack an AGN powerful enough to dilute and/or destroy the PAH
  features.

\end{enumerate}

\acknowledgments
\section{Acknowledgments}
\label{sec:acknowledgments}

We thank the anonymous referee for suggestions which improved this
paper.  Support for MB was provided by the W. M. Keck Foundation.  AHG
acknowledges support from an NSF Small Grant for Exploratory Research
under award AST-0436681.  This work made use of images and data
products provided by the NOAO Deep Wide-Field Survey
\citep{Jannuzi99}, which is supported by the National Optical
Astronomy Observatory (NOAO). NOAO is operated by AURA, Inc., under a
cooperative agreement with the National Science Foundation.  The
Spitzer MIPS and IRAC surveys of the \bootes \ region were obtained
using GTO time provided by the Spitzer Infrared Spectrograph Team (PI:
James Houck), M. Rieke, and the IRAC Team (PI: G. Fazio). IRAC is
supported in part through contract 960541 issued by JPL. The IRS was a
collaborative venture between Cornell University and Ball Aerospace
Corporation funded by NASA through the Jet Propulsion Laboratory and
the Ames Research Center.  Support for this work by the IRS GTO team
at Cornell University was provided by NASA through contract 1257184
issued by JPL/Caltech.

{\it Facilities:} \facility{Spitzer(MIPS, IRAC, IRS)}, \facility{Mayall(Mosaic-1)}, \facility{KPNO:2.1m(ONIS, SQUID, FLAMINGOS, FLAMINGOS-1)}

\bibliographystyle{apj}
\bibliography{references}

\newpage

%%%FIGURE 1
\begin{figure*}
\plotone{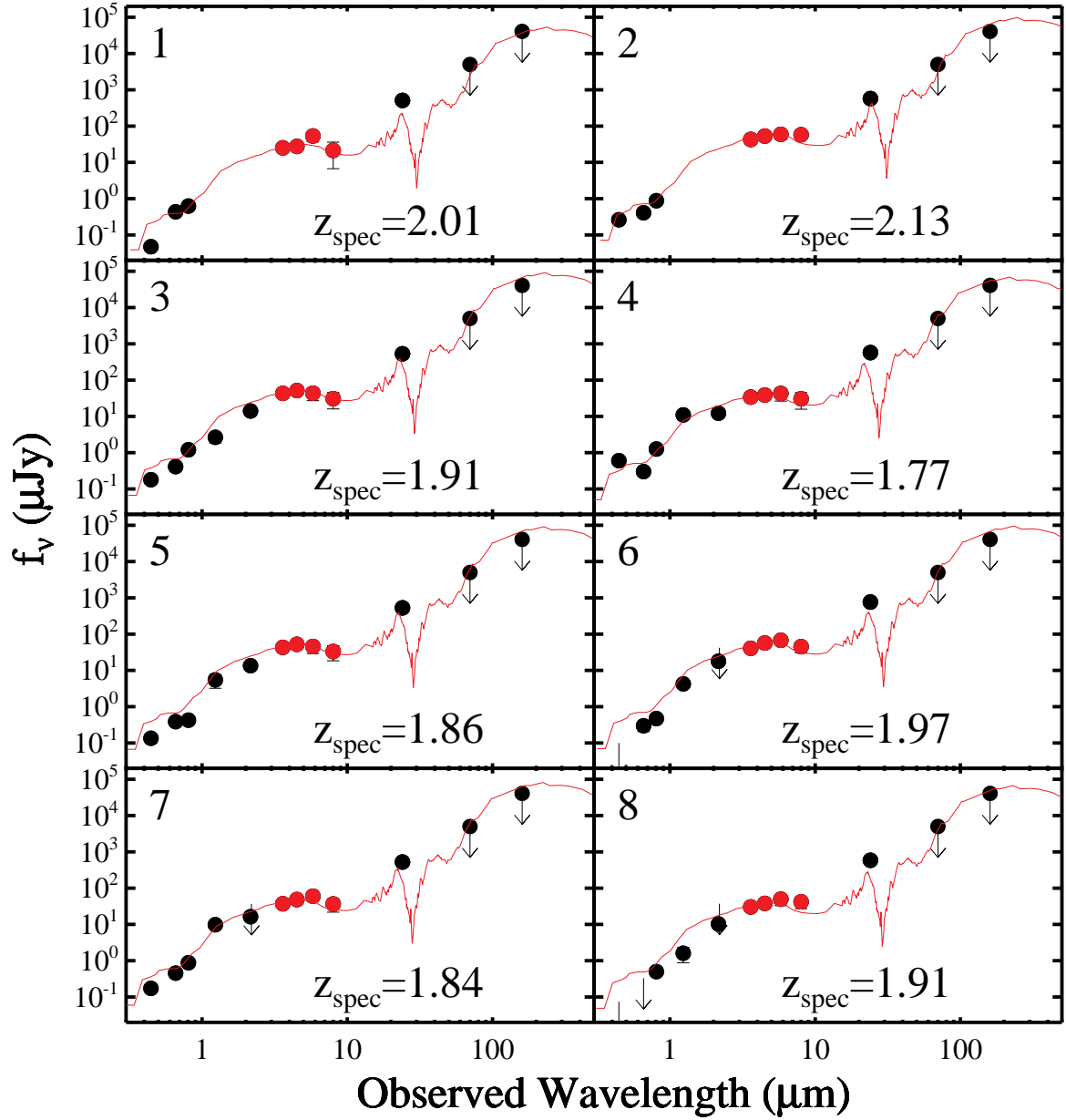}
\caption{The individual SEDs of the bump DOGs are shown as filled
  points.  The red line represents the mid-infrared spectrum of Arp
  220, shifted to the spectroscopic redshift if known (see
  \S{\ref{sec:Redshifts}}) and normalized to best fit the IRAC data
  (red points).  The rest-frame 1.6 $\micron$ stellar photospheric
  bump is clearly seen in the Arp~220 spectrum, and is detected in
  objects at $z \approx 2$ as a turnover in the long-wavelength IRAC
  channels.  All sources in this paper were selected to exhibit this
  turnover.  No color correction has been applied to the
  \textit{Spitzer} photometry.  Points are calibrated to the standard
  spectral shapes adopted for each instrument.}
\label{fig:SEDs}
\end{figure*}

\addtocounter{figure}{-1}
\begin{figure}
  \centering
  \subfigure{
    \includegraphics[width=\textwidth]{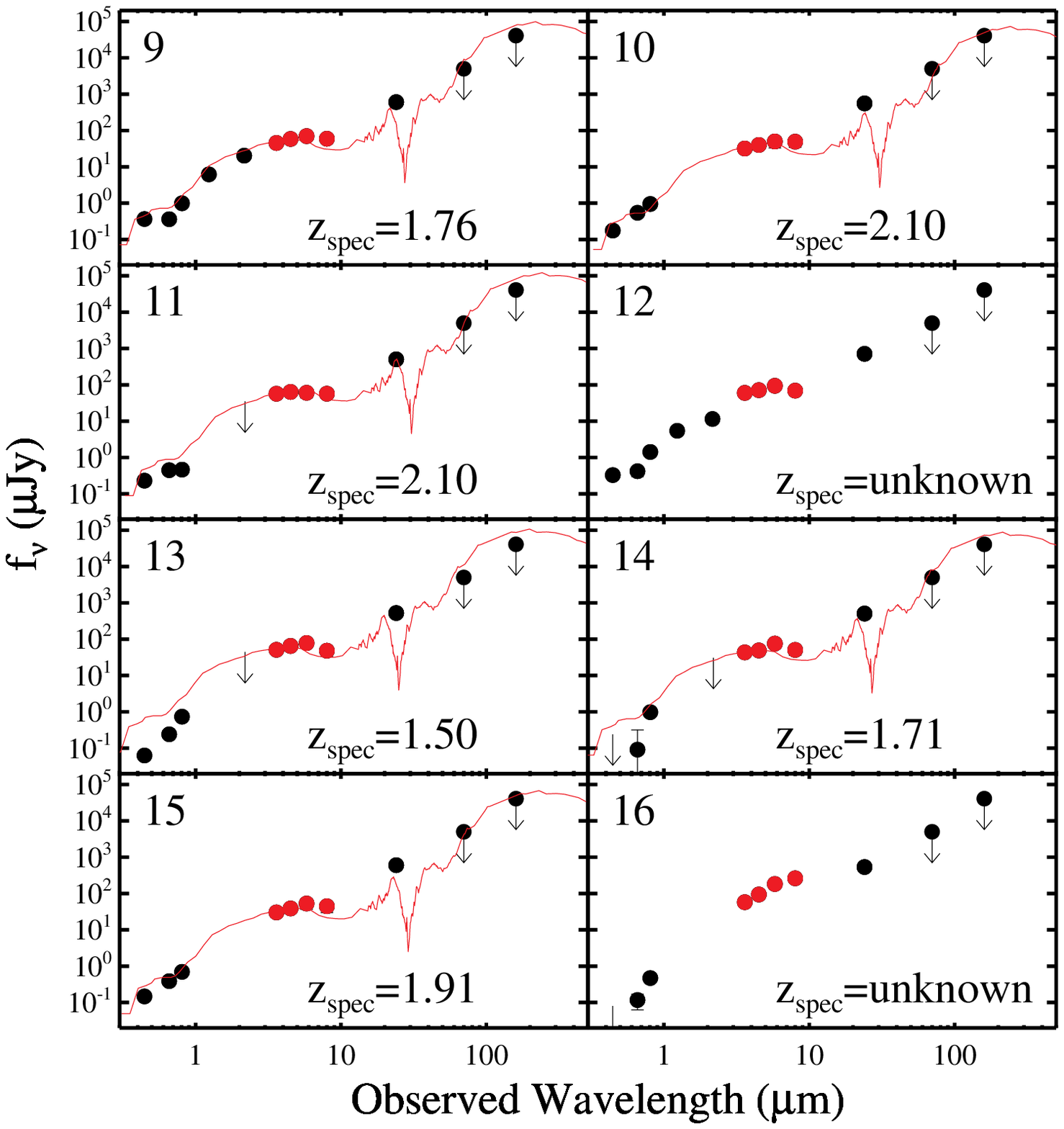}}
  \caption{(cont'd).}
\end{figure}

\addtocounter{figure}{-1}
\begin{figure}
  \centering
  \subfigure{
    \includegraphics[width=\textwidth]{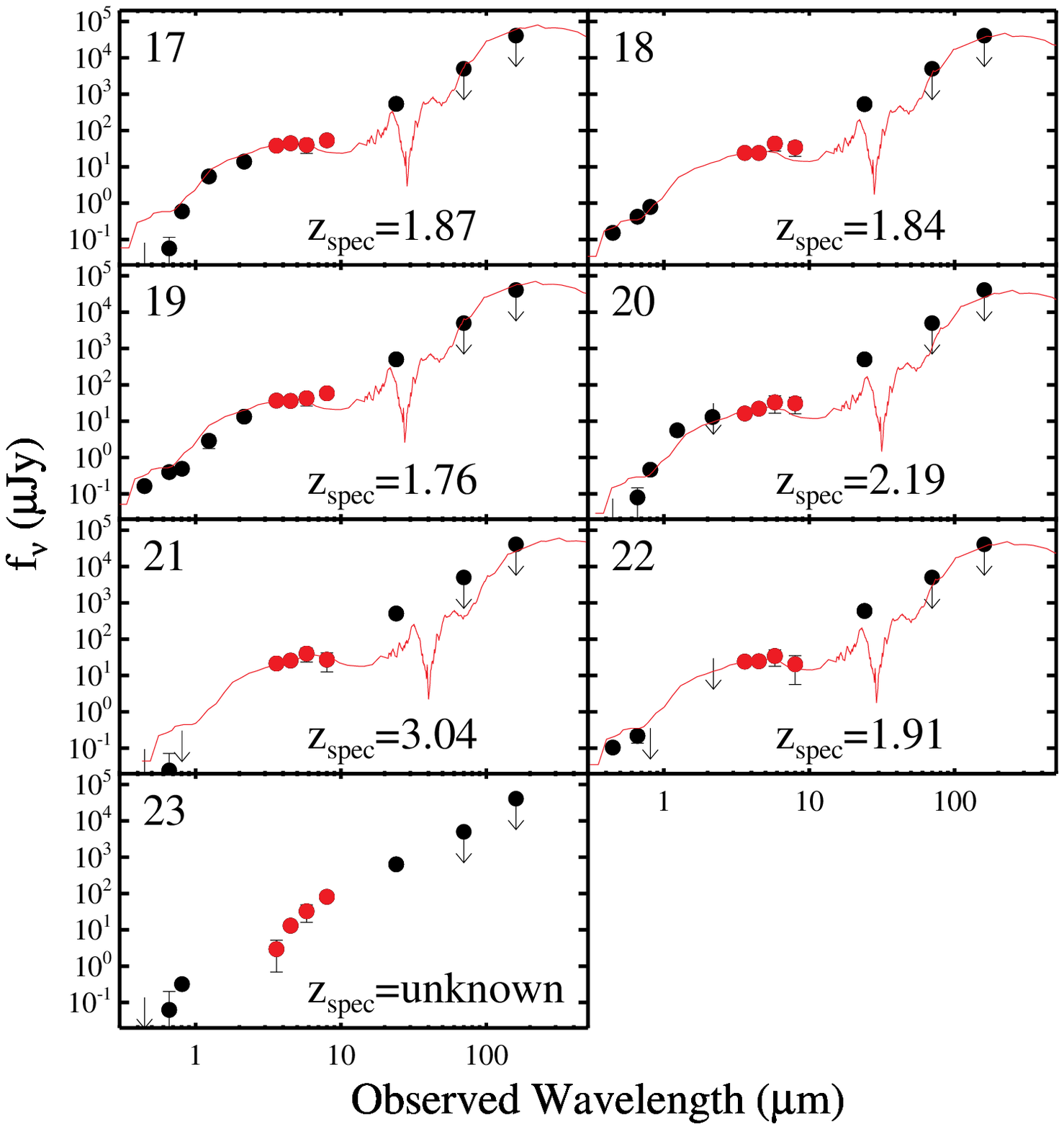}}
  \caption{(cont'd).}
\end{figure}

%%%FIGURE 2
\begin{figure}
\plotone{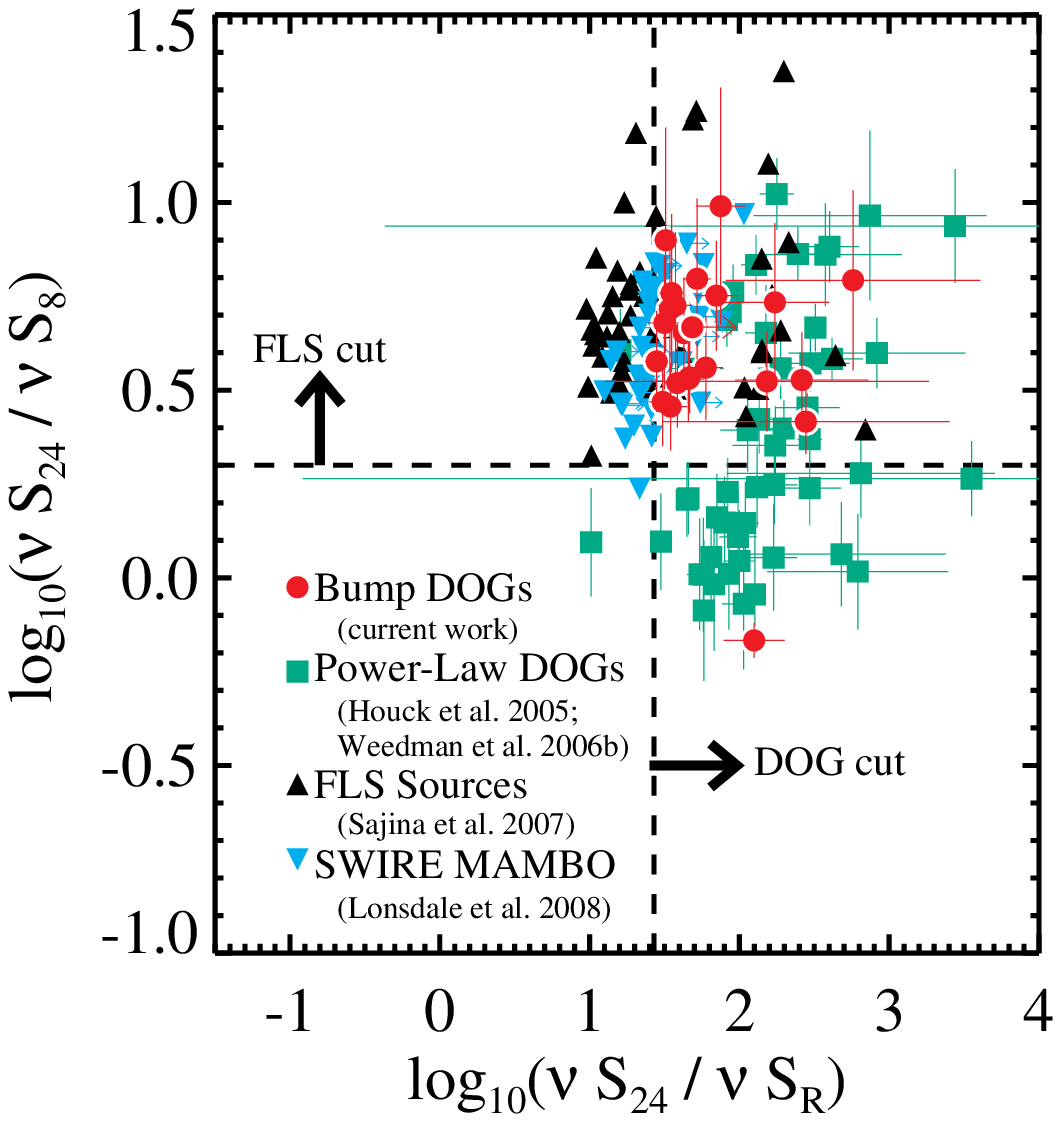}
\caption{Color-color diagram for $z \approx 2$ bump DOGs (red
points), the $f_{\nu}(24~\micron)~>~0.75$~mJy DOG sample published in
\citet{Houck05} and \citet{Weedman06} (green points), and the $z
\approx 2$ ULIRG sample from \citet{Sajina07} (black points).  The
vertical dashed line shows the DOG color criterion
($R-[24]>14$~Vega~mag or $\log_{10}(\nu S_{24} / \nu S_{R}) > 1.43$)
and the horizontal dashed line shows the color criterion used by
\citet{Brand06} to roughly divide starburst and AGN-dominated sources
at $z > 0.6$.  Bump DOGs tend to have larger values of $\log_{10}(\nu
S_{24} / \nu S_{8})$ than their power-law counterparts.}

\label{fig:yancolors}
\end{figure}

%%%FIGURE 3
\begin{figure}
\plotone{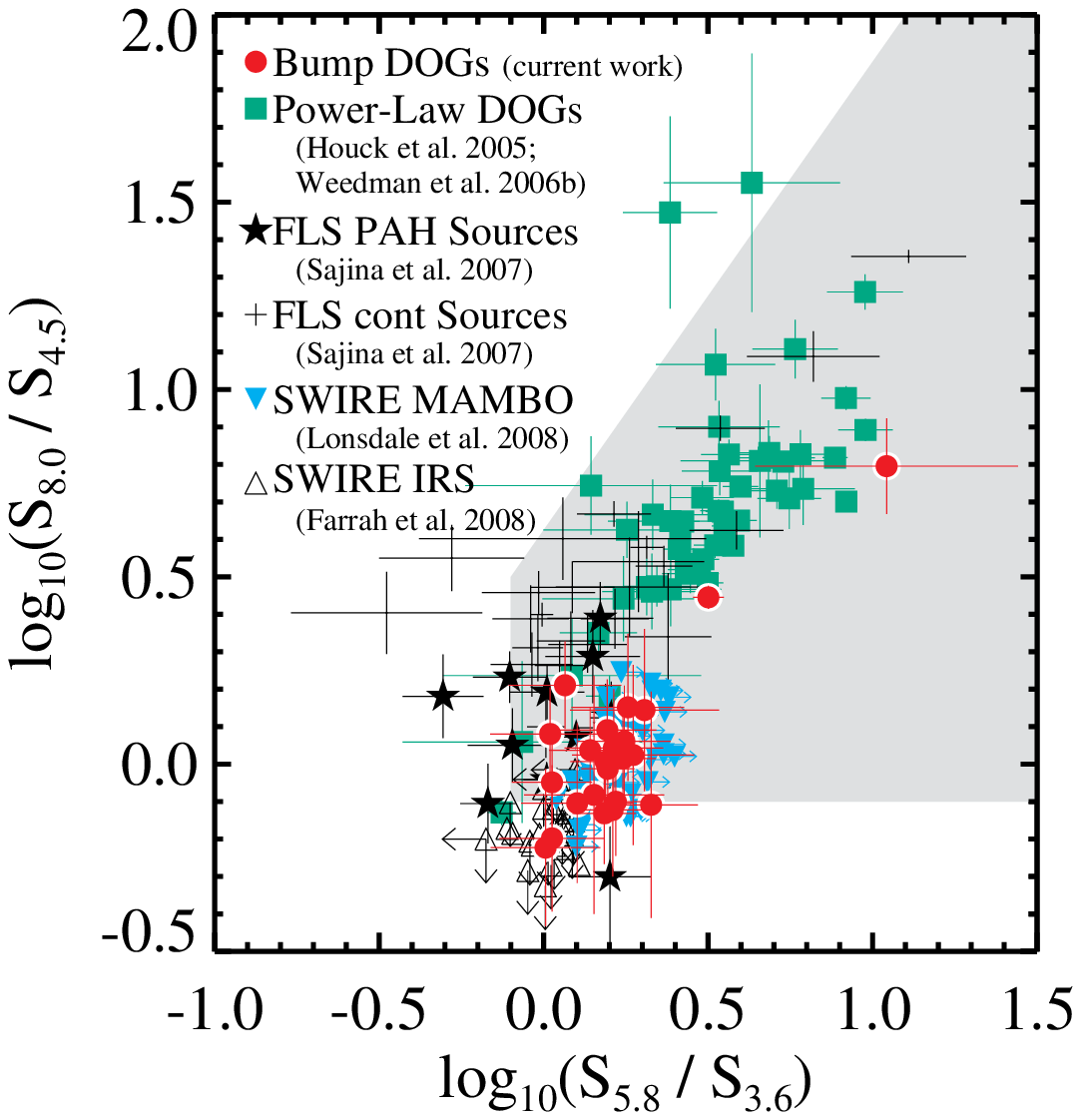}
\caption{IRAC color-color diagram for $z \approx 2$ bump DOGs (red
points), the $f_{\nu}(24~\micron)~>~0.75$~mJy DOG sample published in
\citet{Houck05} and \citet{Weedman06} (green points), and the $z
\approx 2$ ULIRG sample from \citet{Sajina07} (black points).  This
last sample is divided into PAH-rich sources (black stars) and
PAH-poor sources (error bars only).  The wedge used by \citet{Lacy04}
to select AGN-dominated sources is shaded in grey.  Bump sources tend
to lie in a localized region of IRAC color-color space.}
\label{fig:irac}
\end{figure}

%%%FIGURE 4
\begin{figure*}
  \subfigure{
    \includegraphics[width=\textwidth]{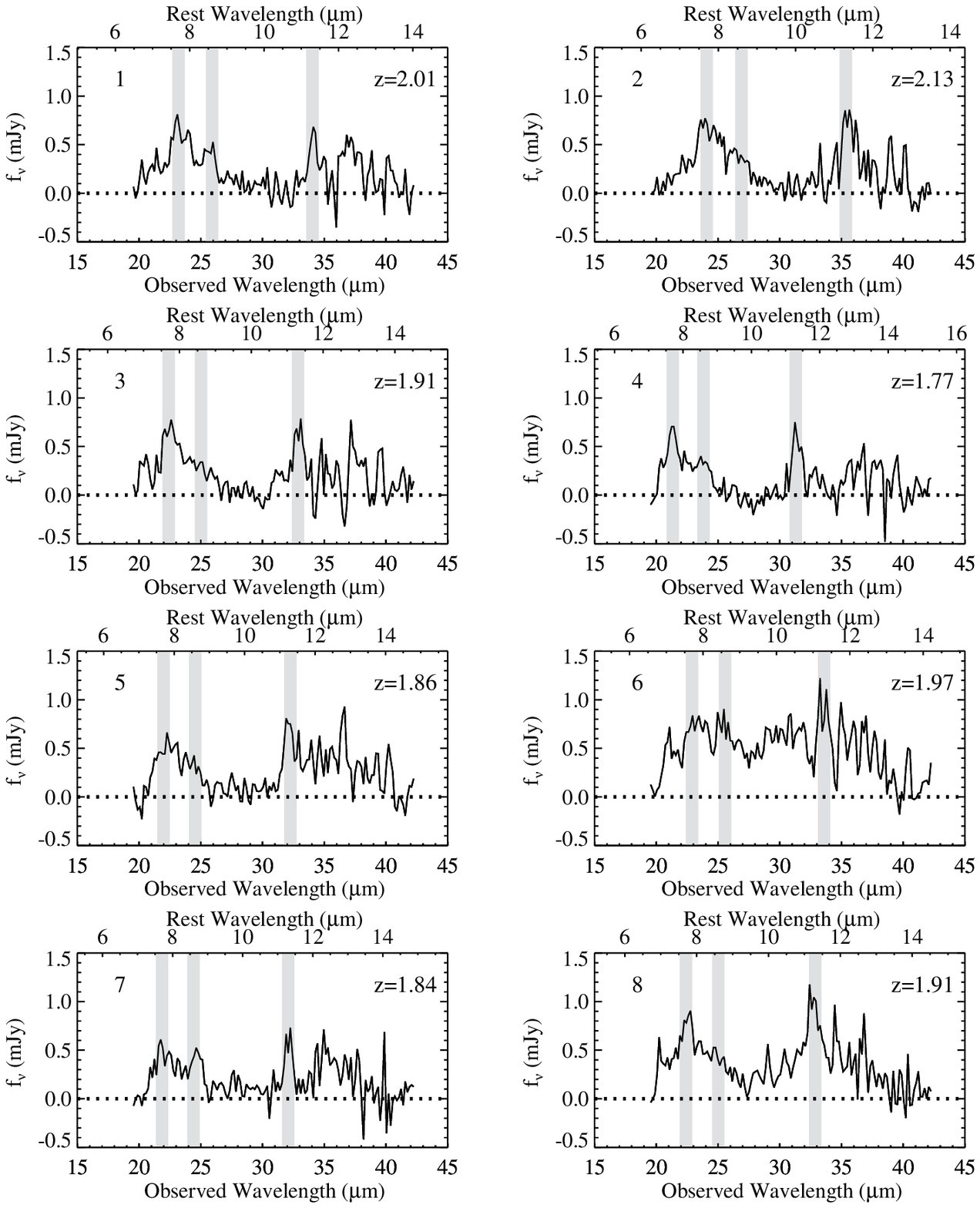}}
  \caption{IRS spectra of $z \approx 2$ bump DOGs.  The shaded
    regions indicate the wavelengths of the 7.7, 8.6, and 11.3
    $\micron$ PAH features at the spectroscopic redshift of the
    source, whether the lines are detected or not.  The vast majority
    of the spectra show two PAH features, consistent with these
    sources being powered by star formation.}
  \label{fig:individualspectra}
\end{figure*}

\addtocounter{figure}{-1}
\begin{figure}
  \centering
  \subfigure{
    \includegraphics[width=\textwidth]{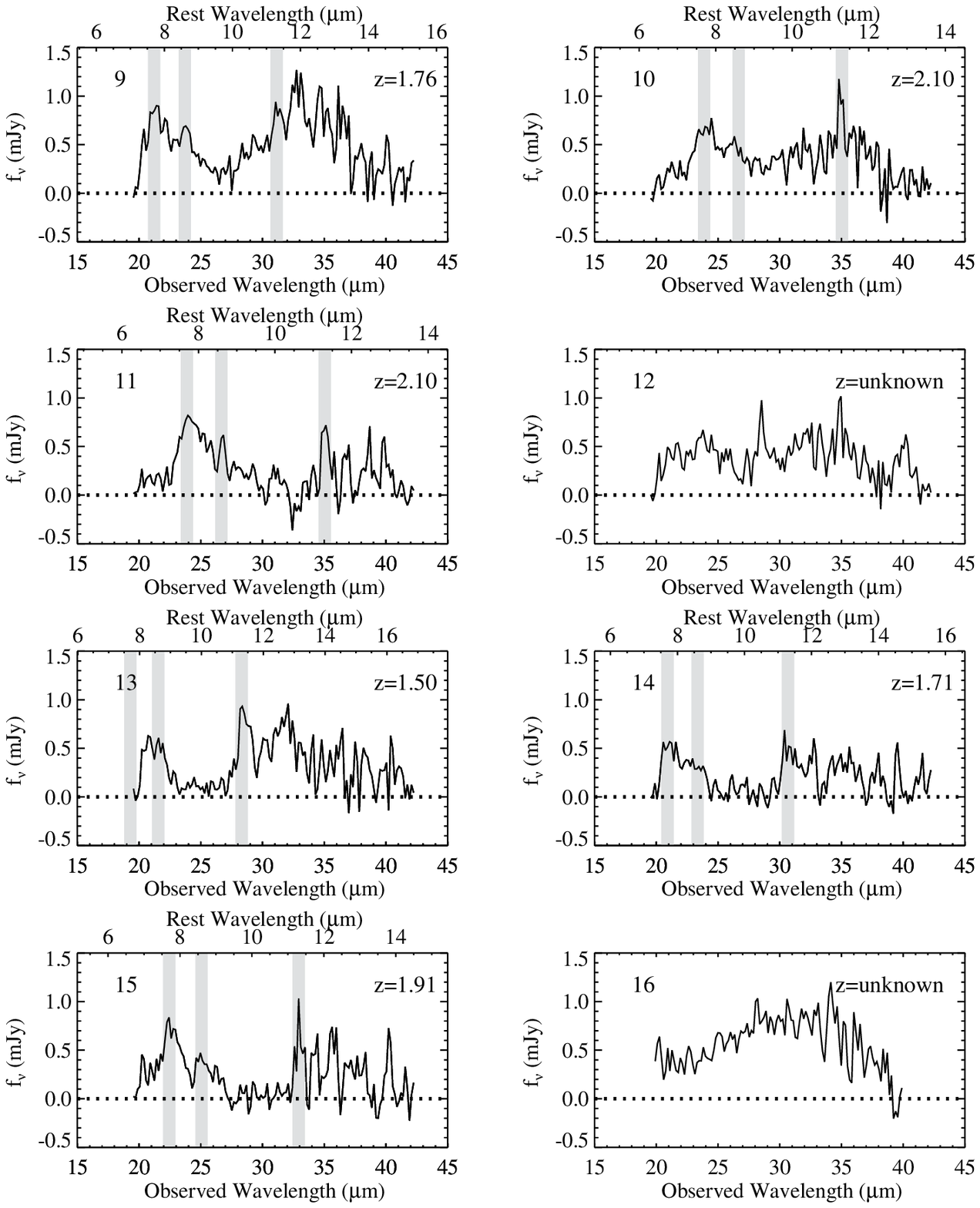}}
  \caption{(cont'd).}
\end{figure}

\addtocounter{figure}{-1}
\begin{figure}
  \centering
  \subfigure{
    \includegraphics[width=\textwidth]{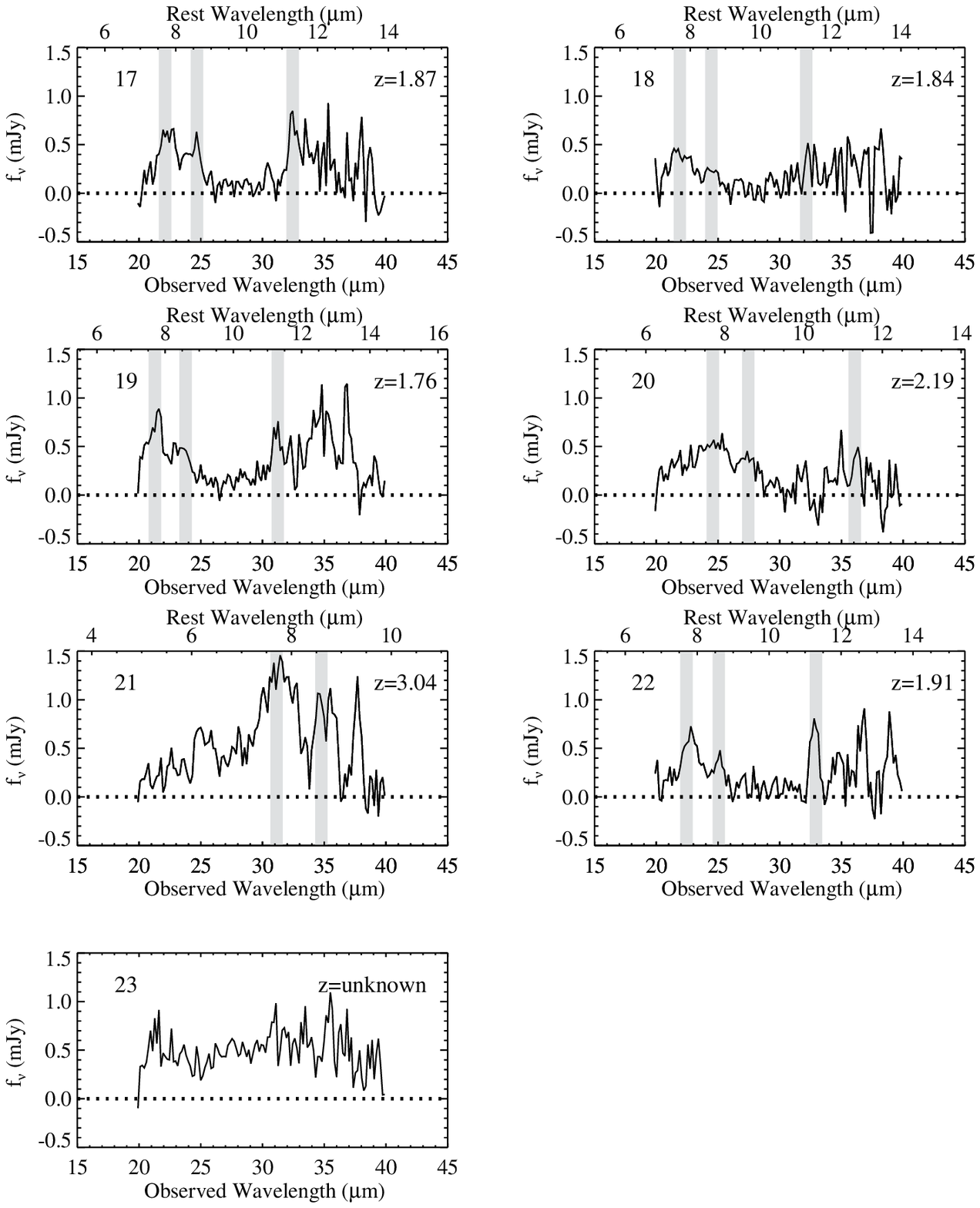}}
  \caption{(cont'd).}
\end{figure}

%%%FIGURE 5
\begin{figure}
\plotone{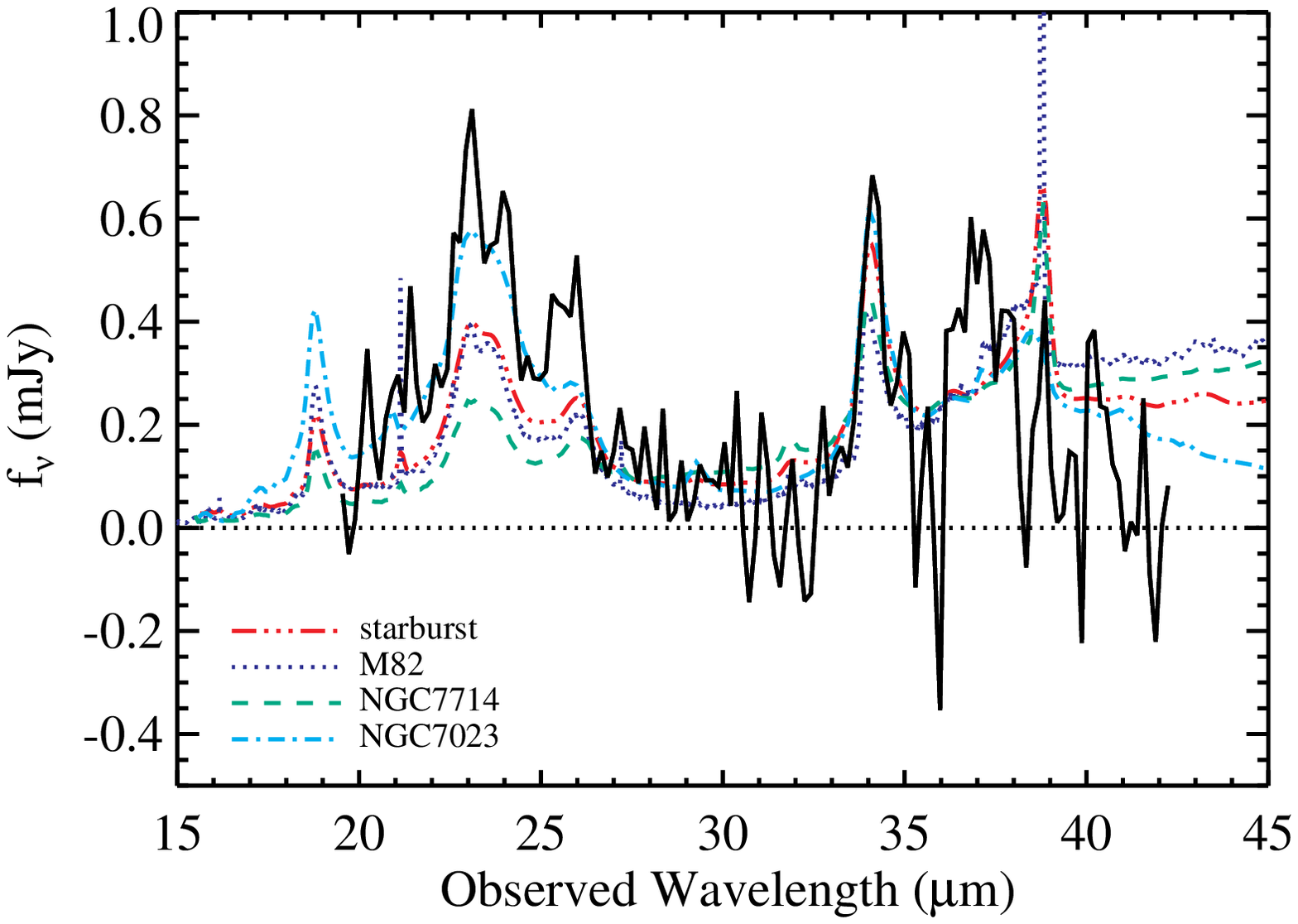}
\caption{Example of redshift determination.  Each observed spectrum,
in this case object 1 (shown in black), is fitted with four template
spectra: the average starburst template from \citet{Brandl06} (red);
M82 (dark blue); NGC 7714 (green); and NGC 7023 (light blue).  The
redshift associated with the best-fitting template is adopted and the
error is taken to be the range of redshifts found for all four
templates.}
\label{fig:examplez}
\end{figure}

%%%FIGURE 6
\begin{figure}
\plotone{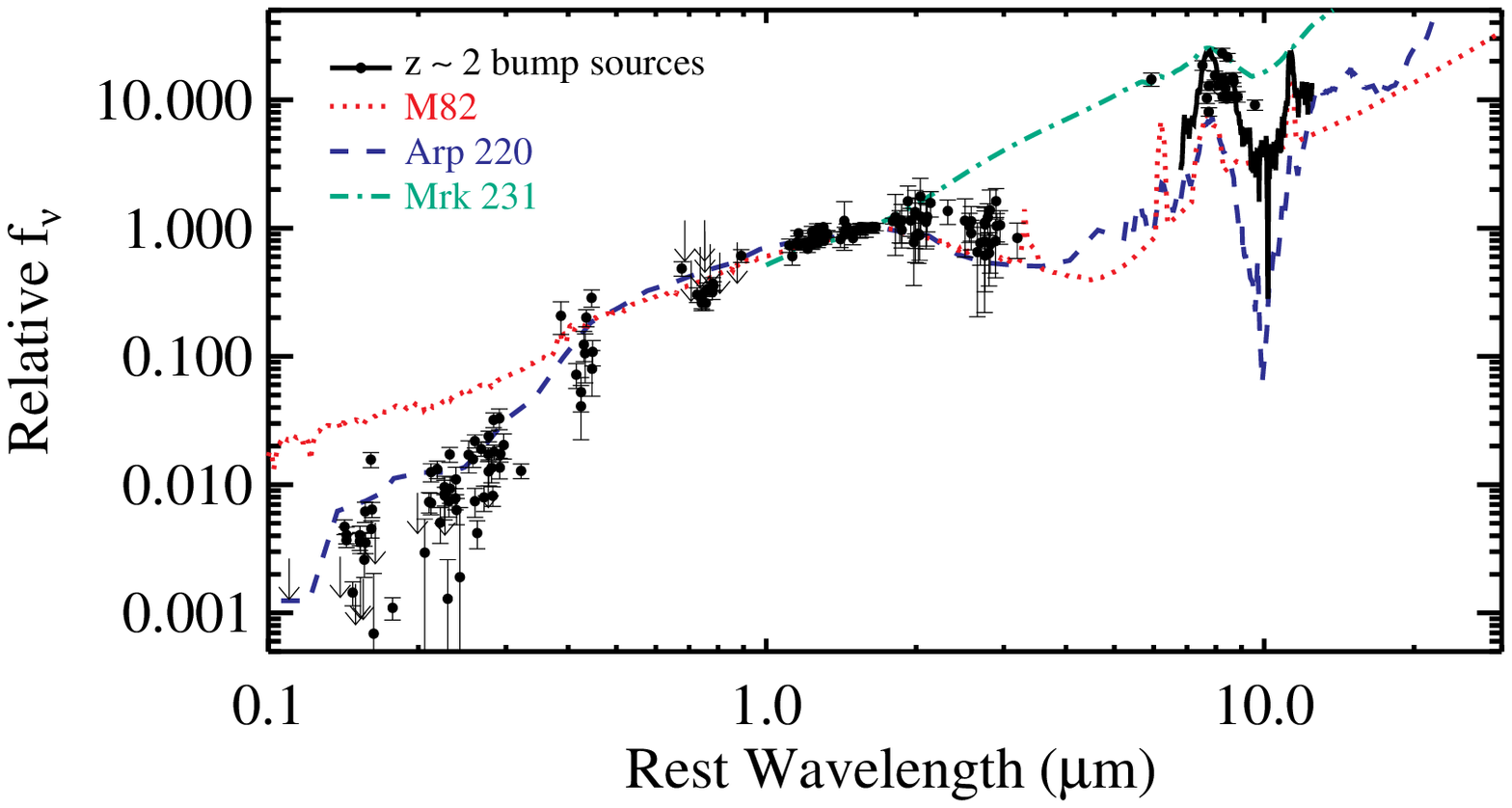}
\caption{Optical to mid-infrared SEDs of $z \approx 2$ bump DOGs.
Each SED is arbitrarily scaled to have a 1.6 $\micron$ flux density of
unity.  The SEDs of bump DOGs strongly resemble that of Arp~220.
As in Figure \ref{fig:SEDs}, no color correction has been applied to
the \textit{Spitzer} photometry.}
\label{fig:mediansed}
\end{figure}

%%%FIGURE 7
\begin{figure}
\plotone{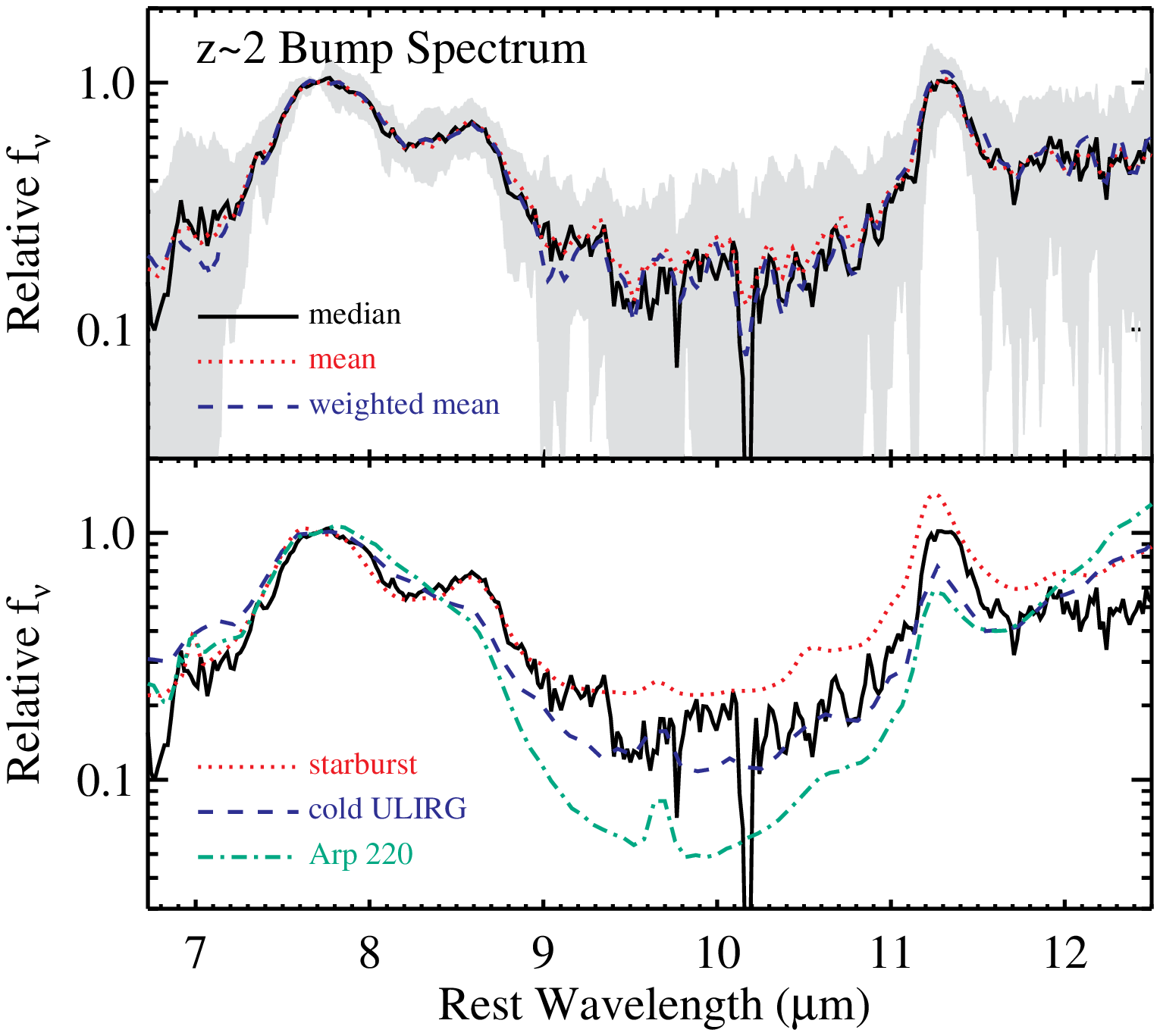}
\caption{\textit{Top:} Median spectrum of 20 $z \approx 2$ bump
DOGs.  The spectrum is normalized to unity at 7.7 $\micron$.
\textit{Bottom:} Same as top panel, but with comparison spectra (also
normalized at 7.7 $\micron$) overplotted.  The mid-infrared spectrum
of bump DOGs appears intermediate between those of local
starbursts and local cold ULIRGs.}
\label{fig:medianspectrum}
\end{figure}

%%%FIGURE 8
\begin{figure}
\epsscale{0.7}
\plotone{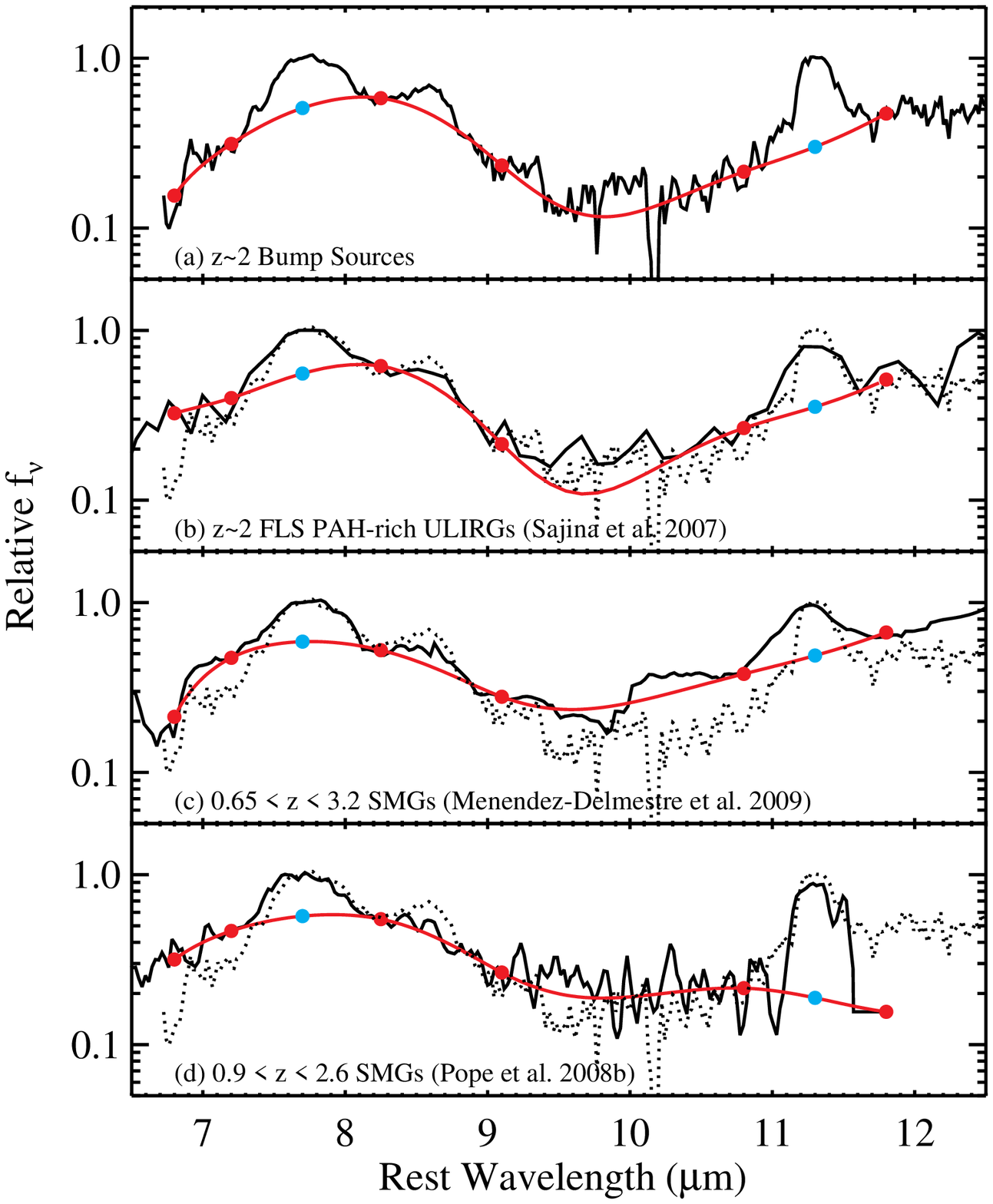}

\caption{Illustration of PAH equivalent width measurements and
  comparison to other high-redshift samples.  Panel (a) shows the
  median spectrum of the bump DOGs (black line).  To determine the
  equivalent width of the PAH features in this spectrum, we choose
  continuum wavelengths of 6.8, 7.2, 8.25, 9.1, 10.8, and
  11.8~$\micron$ (red points).  A spline interpolation was used to
  determine the level of the continuum between these wavelengths (red
  line).  The equivalent width was taken as the integrated line flux
  above this continuum divided by the flux density of the continuum at
  the line center (blue points).  Panels (b) though (d) show selected
  comparison samples (black line) with the median bump DOG
  spectrum overplotted as a dotted line.  The comparison samples have
  composite spectra that are very similar to the bump DOG composite.
  However, the SMG spectrum in panel (c) shows a stronger power-law
  component, which could represent dust heated either by star
  formation or AGN activity.}

\label{fig:measureew}
\end{figure}

\clearpage

%%%FIGURE 9
\begin{figure}
\epsscale{0.7}
\plotone{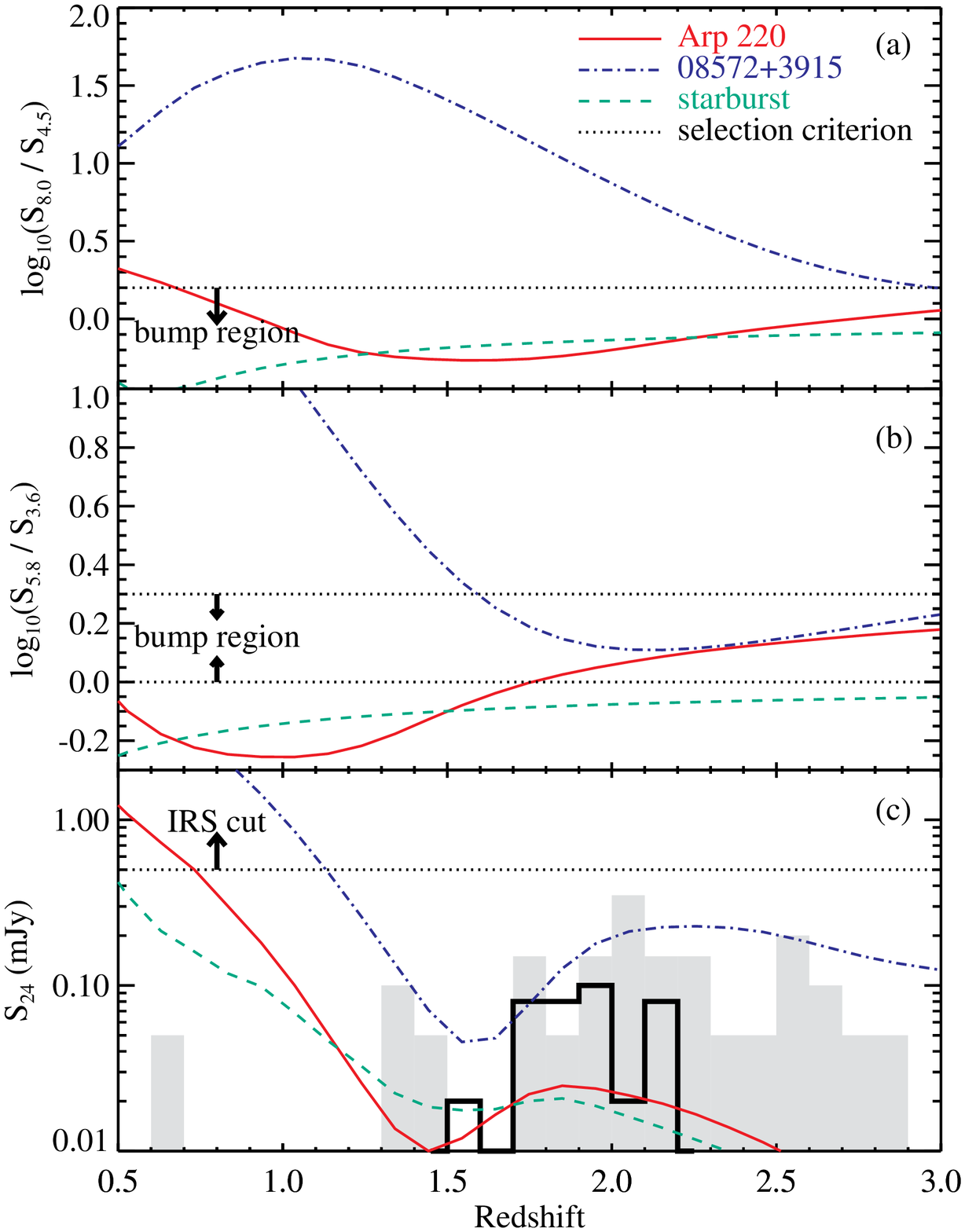}

\caption{Illustration of selection effects.  \textit{Panels (a) and (b):}
  Expected IRAC colors for three local template sources if they were
  placed at various redshifts.  \textit{Panel (c):} The expected 24
  $\micron$ flux density of three local template sources as a function
  of redshift.  The shaded histogram shows the redshift distribution
  of the bump DOGs presented in this paper.  The open histogram shows
  the redshift distribution of the power-law DOGs from \citet{Houck05}
  and \citet{Weedman06}.  While the IRAC selection allows for a
  greater range of redshifts than we observe, the MIPS bump DOG cut
  selects strongly for sources at $z \approx 1.9$.  This is because
  bump DOGs tend to have strong PAH features, which enter the 24
  $\micron$ bandpass at $z \approx 1.9$.  The redshift distribution of
  the power-law sources is more broad because these sources do not
  have strong PAH features.}

\label{fig:selection}
\end{figure}

%%%FIGURE 10
\begin{figure}
\epsscale{0.7}
\plotone{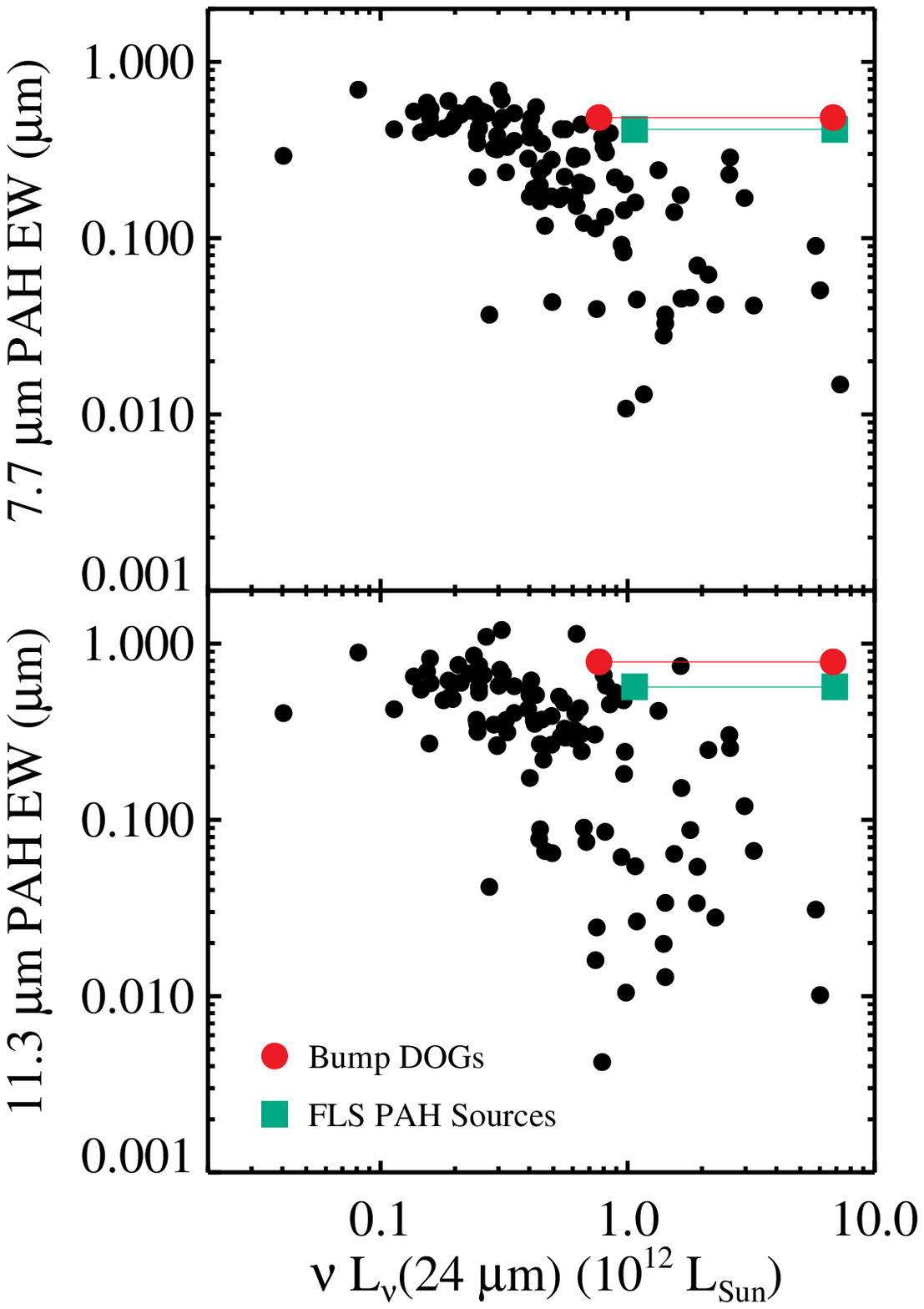}

\caption{PAH equivalent width versus rest-frame 24 $\micron$
luminosity.  The black points represent the 107 ULIRGs from
\citet{Desai07b}.  The green squares represent the average PAH-rich
ULIRG spectrum from \citet{Sajina07}, and the red circles represent
the median spectrum of the $z \approx 2$ bump DOGs presented in
Figure \ref{fig:medianspectrum}.  The endpoints show the minimum and
maximum luminosities of the sources that went into the composite
spectrum.  All equivalent widths were computed using the method
illustrated in Figure \ref{fig:measureew}.  Compared to local ULIRGs
with similar rest-frame 24 $\micron$ luminosities, the bump DOGs
have among the highest 7.7 $\micron$ PAH equivalent widths.}

\label{fig:comparison}
\end{figure}

\clearpage

\clearpage
% Table 1
\begin{deluxetable}{cccccccc}
\tablecaption{The Sample}
%\tabletypesize{\scriptsize}
\tablewidth{0pt}
\startdata \hline\hline
Num     & MIPS Name      & z   & $\Delta z$ & template\tablenotemark{a} \\
\hline

           1 & SST24 J142920.1+333023.9 & 2.01 & 0.02 & NGC 7023 \\ 
           2 & SST24 J143458.9+333437.0 & 2.13 & 0.02 & NGC 7023 \\ 
           3 & SST24 J143324.3+334239.5 & 1.91 & 0.03 & NGC 7023 \\ 
           4 & SST24 J143137.1+334501.6 & 1.77 & 0.02 & NGC 7023 \\ 
           5 & SST24 J143349.6+334601.7 & 1.86 & 0.01 & starburst \\ 
           6 & SST24 J143503.2+340243.6 & 1.97 & 0.02 & NGC 7714 \\ 
           7 & SST24 J142832.4+340849.8 & 1.84 & 0.02 & NGC 7023 \\ 
           8 & SST24 J142941.1+340915.7 & 1.91 & 0.03 & starburst \\ 
           9 & SST24 J142951.2+342042.1 & 1.76 & 0.01 & NGC 7714 \\ 
          10 & SST24 J143321.8+342502.0 & 2.10 & 0.02 & NGC 7714 \\ 
          11 & SST24 J143502.9+342658.8 & 2.10 & 0.02 & NGC 7023 \\ 
          12 & SST24 J142600.6+343452.8 & \nodata & \nodata & \nodata \\ 
          13 & SST24 J143152.4+350030.1 & 1.50 & 0.02 & NGC 7023 \\ 
          14 & SST24 J142724.9+350824.3 & 1.71 & 0.02 & NGC 7023 \\ 
          15 & SST24 J143331.9+352027.2 & 1.91 & 0.02 & NGC 7023 \\ 
          16 & SST24 J143143.4+324943.8 & \nodata & \nodata & \nodata \\ 
          17 & SST24 J143020.5+330344.2 & 1.87 & 0.02 & NGC 7023 \\ 
          18 & SST24 J143816.6+333700.6 & 1.84 & 0.04 & NGC 7023 \\ 
          19 & SST24 J143216.8+335231.7 & 1.76 & 0.02 & M82 \\ 
          20 & SST24 J143743.3+341049.4 & 2.19 & 0.02 & NGC 7023 \\ 
          21 & SST24 J143702.0+344630.4 & 3.04 & 0.02 & starburst \\ 
          22 & SST24 J142652.5+345506.0 & 1.91 & 0.02 & starburst \\ 
          23 & SST24 J142637.4+333025.7 & \nodata & \nodata & \nodata \\

\enddata
\tablecomments{The MIPS name encodes the RA and Dec (J2000) of the 24~$\micron$ source.}
\tablenotetext{a}{The local template spectrum that provided the best fit to the observed IRS spectrum.  This fit was used to determine the redshift.}

\label{table:targets}
\end{deluxetable}

\clearpage

%Table 2
\begin{landscape}
\begin{deluxetable}{cccccccccccccccc}
\tablecaption{Photometric Properties of IRS Targets}
\tabletypesize{\tiny}
%\rotate
\tablewidth{0pt}
\startdata \hline\hline
Num     & $B_W$ & $R$ & $I$ & $J$\tablenotemark{a} & $K_s$\tablenotemark{a} & $K$\tablenotemark{a} & $f_{\nu}(3.6\micron)$ & $f_{\nu}(4.5\micron)$ & $f_{\nu}(5.8\micron)$ & $f_{\nu}(8\micron)$ & $f_{\nu}(24\micron)$ & $f_{\nu}(20{\rm cm})$ \\
        & (mag) & (mag) & (mag) & (mag) & (mag) & (mag) & ($\mu$Jy) & ($\mu$Jy) & ($\mu$Jy) & ($\mu$Jy) & ($\mu$Jy) & ($\mu$Jy) \\ \hline

      1   &         27.2$\pm$0.2   &         24.6$\pm$0.2   &         24.0$\pm$0.1   &              \nodata   &              \nodata   &              \nodata   &         25.2$\pm$2.5   &         27.5$\pm$3.1   &        53.4$\pm$16.6   &        21.4$\pm$14.7   &           510$\pm$40   &                $<$84  \\  
      2   &         25.4$\pm$0.1   &         24.6$\pm$0.2   &         23.6$\pm$0.1   &              \nodata   &              \nodata   &              \nodata   &         42.9$\pm$2.7   &         53.1$\pm$3.3   &        59.4$\pm$16.7   &        57.8$\pm$14.8   &           573$\pm$51   &                $<$84  \\  
      3   &         25.8$\pm$0.2   &         24.6$\pm$0.2   &         23.3$\pm$0.1   &         21.9$\pm$0.3   &         19.2$\pm$0.2   &              \nodata   &         43.1$\pm$2.7   &         51.5$\pm$3.3   &        43.7$\pm$16.6   &        30.8$\pm$14.8   &           530$\pm$37   &                $<$84  \\  
      4   &         24.4$\pm$0.1   &         25.0$\pm$0.2   &         23.2$\pm$0.2   &         20.4$\pm$0.2   &         19.3$\pm$0.1   &              \nodata   &         34.0$\pm$2.6   &         38.8$\pm$3.2   &        42.9$\pm$16.5   &        30.5$\pm$14.8   &           573$\pm$52   &                $<$84  \\  
      5   &         26.1$\pm$0.3   &         24.7$\pm$0.3   &         24.4$\pm$0.2   &         21.2$\pm$0.4   &         19.2$\pm$0.1   &              \nodata   &         43.1$\pm$2.7   &         52.3$\pm$3.3   &        45.7$\pm$16.6   &        33.2$\pm$14.8   &           529$\pm$37   &           132$\pm$28  \\  
      6   &              $>$26.4   &         25.0$\pm$0.3   &         24.3$\pm$0.2   &         21.4$\pm$0.2   &         18.9$\pm$0.1   &              $>$18.0   &         40.8$\pm$2.7   &         56.9$\pm$3.4   &        67.6$\pm$16.7   &        45.1$\pm$14.8   &           764$\pm$58   &           123$\pm$28  \\  
      7   &         25.8$\pm$0.2   &         24.5$\pm$0.3   &         23.6$\pm$0.1   &         20.5$\pm$0.2   &         19.0$\pm$0.1   &              $>$18.1   &         37.1$\pm$2.6   &         48.5$\pm$3.3   &        60.1$\pm$16.7   &        36.6$\pm$14.8   &           524$\pm$35   &                $<$84  \\  
      8   &              $>$26.7   &              $>$24.9   &         24.2$\pm$0.2   &         22.5$\pm$0.5   &         19.5$\pm$0.1   &              $>$18.1   &         30.4$\pm$2.6   &         38.0$\pm$3.2   &        49.6$\pm$16.6   &        42.0$\pm$14.8   &           586$\pm$40   &           300$\pm$28  \\  
      9   &         25.0$\pm$0.1   &         24.8$\pm$0.3   &         23.5$\pm$0.1   &         21.0$\pm$0.2   &         18.8$\pm$0.1   &              $>17.6$   &         45.4$\pm$2.7   &         58.5$\pm$3.4   &        70.0$\pm$16.7   &        59.4$\pm$14.8   &           603$\pm$36   &           195$\pm$28  \\  
     10   &         25.8$\pm$0.2   &         24.3$\pm$0.2   &         23.5$\pm$0.1   &              \nodata   &              \nodata   &              $>17.9$   &         31.9$\pm$2.6   &         39.9$\pm$3.2   &        49.8$\pm$16.6   &        49.2$\pm$14.8   &           556$\pm$41   &                $<$84  \\  
     11   &         25.5$\pm$0.1   &         24.5$\pm$0.2   &         24.3$\pm$0.3   &              \nodata   &              \nodata   &              $>$18.2   &         56.8$\pm$2.8   &         63.7$\pm$3.4   &        60.2$\pm$16.7   &        56.9$\pm$14.8   &           502$\pm$37   &           265$\pm$28  \\  
     12   &         25.1$\pm$0.1   &         24.6$\pm$0.3   &         23.1$\pm$0.1   &         21.2$\pm$0.3   &         19.4$\pm$0.1   &              $>18.2$   &         60.0$\pm$2.8   &         71.3$\pm$3.5   &        94.0$\pm$16.9   &        69.3$\pm$14.9   &           711$\pm$35   &                $<$84  \\  
     13   &         26.9$\pm$0.2   &         25.2$\pm$0.3   &         23.8$\pm$0.1   &              \nodata   &              \nodata   &              $>$17.9   &         51.1$\pm$2.8   &         65.1$\pm$3.4   &        78.2$\pm$16.8   &        48.1$\pm$14.8   &           524$\pm$48   &                $<$84  \\  
     14   &              $>$25.4   &         26.3$\pm$2.7   &         23.5$\pm$0.2   &              \nodata   &              \nodata   &              $>$18.3   &         43.3$\pm$2.7   &         48.9$\pm$3.3   &        75.5$\pm$16.8   &        50.7$\pm$14.8   &           507$\pm$47   &                $<$84  \\  
     15   &         26.0$\pm$0.2   &         24.7$\pm$0.1   &         23.8$\pm$0.1   &              \nodata   &              \nodata   &              $>17.8$   &         30.2$\pm$2.5   &         38.8$\pm$3.2   &        53.1$\pm$16.6   &        44.7$\pm$14.8   &           601$\pm$48   &                $<$84  \\  
     16   &              $>$26.6   &         26.0$\pm$0.5   &         24.3$\pm$0.2   &              \nodata   &              \nodata   &              \nodata   &         57.7$\pm$2.8   &         94.0$\pm$3.7   &       182.8$\pm$17.6   &       262.0$\pm$15.5   &           535$\pm$48   &                $<$84  \\  
     17   &              $>$26.6   &         26.8$\pm$1.1   &         24.0$\pm$0.3   &         21.2$\pm$0.3   &         19.2$\pm$0.1   &              \nodata   &         38.2$\pm$2.6   &         44.5$\pm$3.3   &        39.9$\pm$16.5   &        53.5$\pm$14.8   &           540$\pm$49   &                $<$84  \\  
     18   &         25.9$\pm$0.2   &         24.6$\pm$0.1   &         23.7$\pm$0.1   &              \nodata   &              \nodata   &              \nodata   &         24.2$\pm$2.5   &         24.0$\pm$3.1   &        43.6$\pm$16.6   &        34.0$\pm$14.8   &           530$\pm$36   &                $<$84  \\  
     19   &         25.9$\pm$0.2   &         24.7$\pm$0.3   &         24.2$\pm$0.2   &         21.9$\pm$0.4   &         19.3$\pm$0.1   &              \nodata   &         36.8$\pm$2.6   &         36.0$\pm$3.2   &        42.7$\pm$16.5   &        58.5$\pm$14.8   &           502$\pm$44   &                $<$84  \\  
     20   &              $>$26.7   &         26.4$\pm$0.9   &         24.3$\pm$0.3   &         21.1$\pm$0.3   &         19.3$\pm$0.1   &              $>$18.3   &         16.3$\pm$2.4   &         22.1$\pm$3.1   &        33.0$\pm$16.5   &        30.8$\pm$14.8   &           501$\pm$43   &                $<$84  \\  
     21   &              $>$26.5   &         27.7$\pm$2.1   &              $>$24.7   &              \nodata   &              \nodata   &              \nodata   &         21.4$\pm$2.5   &         25.8$\pm$3.1   &        40.1$\pm$16.5   &        27.3$\pm$14.7   &           508$\pm$60   &           259$\pm$28  \\  
     22   &         26.4$\pm$0.2   &         25.3$\pm$0.4   &              $>$24.6   &              \nodata   &              \nodata   &              $>$18.3   &         24.2$\pm$2.5   &         24.7$\pm$3.1   &        34.4$\pm$16.5   &        20.4$\pm$14.7   &           598$\pm$50   &                $<$84  \\  
     23   &              $>$26.0   &         26.7$\pm$2.4   &         24.7$\pm$0.3   &              \nodata   &              \nodata   &              \nodata   &          2.9$\pm$2.3   &         13.0$\pm$3.0   &        32.5$\pm$16.5   &        81.3$\pm$14.9   &           636$\pm$49   &                $<$84  \\

\enddata
\tablecomments{All magnitudes are on the Vega system.  All limits are 3-$\sigma$ limits.}
\tablecomments{Source 1 through 15 were also the targets of pointed \textit{Spitzer}/MIPS observations, all of which led to 3$\sigma$ upper limits of 5000 $\mu$Jy and 40,800 $\mu$Jy at 70 and 160 $\micron$, respectively.  See \S{\ref{sec:MIPS}}.}
\tablenotetext{a}{The ellipses indicate that the source was not covered by the observations.}
\label{table:photometry}
\end{deluxetable}
\clearpage
\end{landscape}

%Table 3
\begin{deluxetable}{cccccccccccccccc}
\tablecaption{Luminosities}
%\tabletypesize{\scriptsize}
\tablewidth{0pt}
\startdata \hline\hline
Num     & $\nu {\rm L}_{\nu}(24 \micron)$ & ${\rm L}_{8-1000 \micron}$ (IRS) & ${\rm L}_{8-1000 \micron}$ (radio,SMG)\tablenotemark{a} & ${\rm L}_{8-1000 \micron}$ (radio,local)\tablenotemark{b} \\
        & $10^{12} {\rm L}_{\odot}$        & $10^{12} {\rm L}_{\odot}$                   & $10^{12} {\rm L}_{\odot}$                  & $10^{12} {\rm L}_{\odot}$ \\ \hline

           1 & 0.98 & 7.64 & $<$3.01 & $<$11.44 \\ 
           2 & 1.30 & 10.17 & $<$3.42 & $<$13.01 \\ 
           3 & 0.88 & 6.86 & $<$2.66 & $<$10.12 \\ 
           4 & 0.87 & 6.79 & $<$2.23 & $<$8.48 \\ 
           5 & 0.83 & 6.44 & 3.91 & 14.87 \\ 
           6 & 1.39 & 10.82 & 4.21 & 15.99 \\ 
           7 & 0.81 & 6.29 & $<$2.44 & $<$9.27 \\ 
           8 & 0.97 & 7.58 & 9.50 & 36.13 \\ 
           9 & 0.92 & 7.21 & 5.09 & 19.34 \\ 
          10 & 1.22 & 9.48 & $<$3.32 & $<$12.63 \\ 
          11 & 1.10 & 8.56 & 10.48 & 39.84 \\ 
          12 & \nodata & \nodata & \nodata & \nodata \\ 
          13 & 1.04 & 8.13 & $<$1.53 & $<$5.81 \\ 
          14 & 0.81 & 6.31 & $<$2.07 & $<$7.87 \\ 
          15 & 1.01 & 7.83 & $<$2.68 & $<$10.18 \\ 
          16 & \nodata & \nodata & \nodata & \nodata \\ 
          17 & 0.86 & 6.68 & $<$2.54 & $<$9.65 \\ 
          18 & 0.82 & 6.39 & $<$2.45 & $<$9.33 \\ 
          19 & 0.77 & 5.96 & $<$2.21 & $<$8.42 \\ 
          20 & 1.27 & 9.87 & $<$3.66 & $<$13.92 \\ 
          21 & 6.79 & 52.92 & 23.85 & 90.66 \\ 
          22 & 1.00 & 7.80 & $<$2.68 & $<$10.18 \\ 
          23 & \nodata & \nodata & \nodata & \nodata \\

\enddata
\tablenotetext{a}{These values were calculated from the observed radio fluxes assuming the FIR-radio correlation calibrated on SMGs.}

\tablenotetext{b}{These values were calculated from the observed radio fluxes assuming the FIR-radio correlation calibrated on nearby galaxies.}

\label{table:luminosities}
\end{deluxetable}

\end{document}